\documentclass[12pt]{article}
\pdfoutput=1
\usepackage{epsf,color,colordvi}
\usepackage{amsmath}
\usepackage{verbatim}
\usepackage{amssymb}
\usepackage[most]{tcolorbox}
\usepackage{cite}
\usepackage{fontenc}
\usepackage{graphicx}
\usepackage{float}
\usepackage[lofdepth,lotdepth,caption=false]{subfig}
\usepackage{color}
\usepackage{booktabs}
\usepackage{tabularx}
\usepackage{multirow}
\usepackage{longtable}
\usepackage{lscape}
\usepackage[font=small,skip = 0pt]{caption}
\usepackage{dcolumn}
\usepackage{rotating}
\usepackage{graphics}
\usepackage[utf8x]{inputenc}
\usepackage{hyperref}
\usepackage[normalem]{ulem}
\usepackage{chngcntr}
\definecolor{darkgreen}{cmyk}{1,0,1,0.4}
\definecolor{brown}{cmyk}{0,0.8,1,0.2}
\definecolor{darkred}{cmyk}{0,1,1,0.2}

\allowdisplaybreaks[4] 
\makeatletter
\renewcommand{\fnum@table}{\textbf{\tablename~\thetable}}
\renewcommand{\fnum@figure}{\textbf{\figurename~\thefigure}}
\makeatother

\newcounter{myenumi}

\renewcommand{\themyenumi}{\roman{myenumi}}
{\end{list}}

\newlength{\myem}
\newcounter{mysubequation}[equation]

\makeatletter
\renewcommand{\section}{\@startsection{section}{1}{0em}{-\baselineskip}%
{\baselineskip}{\normalfont\large\bfseries}}
\renewcommand{\subsection}%
{\@startsection{subsection}{2}{0em}{-0.7\baselineskip}%
{0.7\baselineskip}{\normalfont\bfseries}}
\makeatother

\newcommand{\bi}{\begin{itemize}}
\newcommand{\ei}{\end{itemize}}

\newcommand{\bea}{\begin{eqnarray}}
\newcommand{\eea}{\end{eqnarray}}
\def\beq{\begin{equation}}
\def\eeq{\end{equation}}

\newcommand{\td}{\theta_{14}}

\newcommand{\ldm}{\Delta m_{31}^2}
\newcommand{\sdm}{\Delta m_{21}^2}
\newcommand{\lldm}{\Delta m_{41}^2}
\newcommand{\dn}{\delta[\Delta N^{CP}_{\alpha \beta}]}

\newcommand{\ie}{{\it i.e.}}

\newcommand{\chisq}{\Delta\chi^2}

\def\<{\langle}
\def\>{\rangle}

\def\dfrac#1#2{{\displaystyle\frac{#1}{#2}}}

\def\lsim{\mathrel{\rlap{\lower4pt\hbox{\hskip1pt$\sim$}}
    \raise1pt\hbox{$<$}}}         
\def\gsim{\mathrel{\rlap{\lower4pt\hbox{\hskip1pt$\sim$}}
    \raise1pt\hbox{$>$}}}         

\newcommand{\dacp}[1]{\ensuremath{\delta [\Delta P^{CP/T}_{\alpha\beta}]}}

\begin{document}
\begin{titlepage}

\vspace*{-3.cm}
\begin{flushright}

\end{flushright}


\renewcommand{\thefootnote}{\fnsymbol{footnote}}
\setcounter{footnote}{-1}

{\begin{center}
{\large\bf 
Signals of eV-scale sterile neutrino at long baseline neutrino experiments 
\\[0.2cm]
}
\end{center}}

\renewcommand{\thefootnote}{\alph{footnote}}

\vspace*{.8cm}
\vspace*{.3cm}
{
\begin{center} 

   {\sf                 Sabila Parveen$^{\S}$\,  \footnote[1]{\makebox[1.cm]{Email:} sabila41\_sps@jnu.ac.in},
                }
        {\sf               Kiran Sharma$^{\P}$\,  \footnote[2]{\makebox[1.cm]{Email:} kirans@iitbhilai.ac.in}
        }
            {\sf               Sudhanwa Patra$^{\P}$\,  \footnote[3]{\makebox[1.cm]{Email:} sudhanwa@iitbhilai.ac.in}        
                }                
and                 
            {\sf                 Poonam Mehta$^{\S}$\,\footnote[4]{\makebox[1.cm]{Email:} pm@jnu.ac.in}
}
\end{center}
}
\vspace*{0cm}
{\it 
\begin{center}
$^\S$\, School of Physical Sciences, Jawaharlal Nehru University, 
      New Delhi 110067, India  \\
$^\P$\, Department of Physics,  Indian Institute of Technology Bhilai 492013, India   \\
\end{center}
}

{\Large 
\bf
 \begin{center} Abstract  
\end{center}}
While most of the results of the neutrino oscillation experiments can be accommodated within the standard paradigm of three active flavor, there are tantalizing hints of an light eV-scale sterile neutrino from anomalous results of a few short baseline experiments. This additional light sterile neutrino is expected to leave an imprint on the physics observables pertaining to standard unknowns such as determination of the Dirac-type leptonic $CP$ phase, $\delta_{13}$, the question of neutrino mass hierarchy and the octant of $\theta_{23}$. The upcoming long baseline neutrino experiments such as T2HK, DUNE and P2O will be sensitive to active - sterile mixing. In the present work, we examine and assess the capability of these long baseline experiments to probe the sterile neutrino at the level of probabilities and event rates. We perform a detailed study by taking into account the values of parameters that are presently allowed and (a) study the impact on  $CP$ violation by examining the role played by various appearance and disappearance channels, (b) address the question of disentangling the intrinsic effects from extrinsic effects in the standard paradigm as well as three active plus one light sterile neutrino, and finally (c) assess the ability of these long baseline experiments to distinguish between the two scenarios. Our results indicate that for the true values of sterile parameters and for all values of $\delta_{13}$,  the sensitivity of P2O is the lowest while the sensitivity of T2HK is modest ($<3\,\sigma$) and the sensitivity of DUNE is  $> 3\,\sigma$. For larger values of the sterile mixing angles, there is an improvement in the sensitivity for all the three considered experiments.

\vspace*{.5cm}

\end{titlepage}

\newpage

\renewcommand{\thefootnote}{\arabic{footnote}}
\setcounter{footnote}{0}
\section{Introduction}
\label{sec:introduction}
There is compelling evidence in support of neutrino oscillations (among the three active flavor, $\nu_e, \nu_\mu, \nu_\tau$) from various oscillation experiments. Accommodating the experimental confirmation of neutrino flavor oscillations requires physics beyond the Standard Model (SM). Most of the parameters responsible for oscillations have been measured to a fair degree of precision (via a global-fit~\cite{deSalas:2020pgw, Esteban:2020cvm, Capozzi:2021fjo} of the oscillation data within the framework of three active neutrinos). However, some questions still allude us, such as - whether $CP$ (and $T$)~\footnote{$CP$ refers to charge-conjugation and parity symmetry, $T$ refers to time-reversal symmetry.} is violated, what is the hierarchy of neutrino masses and what is the correct  octant of $\theta_{23}$. Out of nine flavor parameters in the standard three flavor neutrino mixing framework, only six~\footnote{The absolute mass scale and two Majorana phases are not accessible in oscillation experiments.} can be accessed via neutrino oscillation experiments. These are  three angles ($\theta_{12}, \theta_{13}, \theta_{23}$), single Dirac-type $CP$ phase ($\delta_{13}$) and two mass-squared differences ($\Delta m^2_{21}, \Delta m^2_{31}$ with $\Delta m^2_{ij} = m^2_{i}-m^2_{j}$). Of these, the three angles and the Dirac-type $CP$ phase appear in the $3 \times 3$ leptonic mixing matrix, commonly referred to as the Pontecorvo-Maki-Nakagawa-Sakata (PMNS) matrix~\cite{Pontecorvo:1957qd,Pontecorvo:1957cp,Gribov:1968kq,Maki:1962mu}. In vacuum, $\delta_{13}$ induces $CP$ and $T$ violation effects and we shall refer to these as {\bf{intrinsic}} effects~\footnote{Matter induces additional $CP$ violating effects (as matter is $CP$ asymmetric) and this makes it very hard to measure the value of the $CP$ phase that appears in the mixing matrix~\cite{Bilenky:1997dd, Nunokawa:2007qh, Rout:2017udo}. These effects are referred to as {\bf{extrinsic or fake}} as these obscure the determination of intrinsic $CP$ phase ($\delta_{13}$).}.

The goal of current and future oscillation experiments is to measure the known parameters with improved precision and to measure the unknown parameters. 
{{Some of the upcoming long baseline experiments at different baselines such as Tokai to Hyper-Kamiokande (T2HK) at $295$ Km~\cite{Hyper-Kamiokande:2018ofw}, Deep Underground Neutrino Experiment (DUNE) at $1300$ Km~\cite{Acciarri:2015uup, Abi:2020evt, DUNE:2020fgq} and Protvino to ORCA (P2O) at $2595$ Km~\cite{Akindinov:2019flp, KM3NET:2016zxf} present excellent opportunities to resolve the remaining questions in neutrino oscillation physics.

Although data from various oscillation experiments fits well within the framework of standard three neutrino paradigm, there are anomalous results from some of the short baseline experiments hinting towards the possible existence of light sterile states (see~\cite{Giunti:2019aiy, Dasgupta:2021ies} for comprehensive reviews on the topic). 
The first hint came from the Liquid Scintillator Neutrino Detector (LSND)  where an excess of $\bar{\nu}_{e}$ events was reported  at a significance of $3.8\,\sigma$~\cite{LSND:1996ubh}.The smaller version of Booster Neutrino Experiment (BooNE), MiniBooNE supported the LSND  claim at  $4.8\,\sigma$. The significance of combined LSND and MiniBooNE excesses is found to be $6.1\,\sigma$~\cite{MiniBooNE:2020pnu}. In recent times, the MicroBooNE experiment reignited the debate on existence of sterile neutrinos as they claimed no evidence in favour of sterile neutrinos~\cite{MicroBooNE:2022sdp}. However, an analysis using electron disappearance channel supported the idea of active-sterile oscillation~\cite{Denton:2021czb} (see also~\cite{Arguelles:2021meu}).  
Additionally, a combined fit of MiniBooNE and MicroBooNE  reveals a  preference for oscillation among the three active neutrino states and one sterile neutrino state with high confidence~\cite{MiniBooNE:2022emn}. Moreover, the Gallium-based radio-chemical experiments used to study solar neutrinos such as SAGE~\cite{SAGE:2009eeu}, GALLEX~\cite{Kaether:2010ag} and Baskan Experiment to Sterile transition (BEST)~\cite{Barinov:2021asz}  have reported a deficit of $\nu_e$ events at a significance of  $> 3\,\sigma$ hinting towards existence of light sterile neutrino. Similar support came from the  reactor anti-neutrino anomaly (RAA)~\cite{Huber:2011wv, Mention:2011rk} and several experiments have been planned to test this anomaly. In this connection,  the Neutrino-4 experiment supported the active-sterile oscillation hypothesis with $\lldm \sim 7 \,\text{eV}^{2}$ and $\sin^{2}2\td \sim 0.36$ at $3\,\sigma$ C.L.~\cite{Serebrov:2020kmd}. There is also tension between the appearance ($\nu_\mu \to \nu_e$) and disappearance ($\nu_e \to \nu_e$ and $\nu_\mu \to \nu_\mu$) data sets and the situation is inconclusive presently~\cite{Dentler:2018sju}.  

{{In future, the Short Baseline Neutrino (SBN) program~\cite{Machado:2019oxb}  at Fermilab  is expected to clarify the situation regarding the existence of eV-scale sterile neutrinos. The SBN program comprises of three detectors : 
(a) Short-Baseline Near Detector (SBND) at 110 meters from the Booster neutrino beam (BNB) target and with 112 tons of liquid argon within the active volume of its detection systems~\cite{SBND:2024vgn}, 
(b) MicroBooNE~\cite{MicroBooNE:2022sdp} located at 470 meters from the BNB target. It consists of a 8250-wire TPC and 32 photomultiplier tubes which instrument 80 tons of liquid argon in the active volume, and (c) Short-Baseline Far Detector (SBFD), the 
ICARUS T600~\cite{Torretta:2024fbn} detector located at 600 meters from the BNB target and holds 500 tons of liquid argon in the active volumes. Also, JSNS$^2$ (J-PARC Sterile Neutrino Search at J-PARC Spallation Neutron Source)~\cite{JSNS2:2017gzk} is expected to search for $\bar \nu_\mu \to \bar \nu_e$ oscillation and directly test the LSND claim. 
}}

 Some theoretical and phenomenological consequences of eV-scale sterile neutrino have been widely explored in literature~\cite{Goswami:1995yq, Gandhi:2015xza,  Parke:2015goa, Dutta:2016glq, Kosmas:2017zbh, Agarwalla:2018nlx, Reyimuaji:2019wbn,  Chatterjee:2022pqg, Singha:2022btw, Chattopadhyay:2022hkw,Majhi:2019hdj,Fiza:2021gvq}. For instance, the impact of sterile neutrinos on $CP$ sensitivity at long baseline experiments~\cite{Dutta:2016glq}, resolution of the octant degeneracy~\cite{Chatterjee:2022pqg}, impact of active-sterile mixing in the context of T2HK~\cite{Agarwalla:2018nlx} and other long baseline experiments~\cite{Majhi:2019hdj}, the role of sterile phases~\cite{Fiza:2021gvq}, the role of near and far detector~\cite{DUNE:2020fgq}  and so on. In the present work, we (a) study the impact on $CP$ violation by examining the role played by various appearance and disappearance channels, (b) address the question of disentangling the intrinsic effects from extrinsic effects in the standard paradigm as well as three active plus one light sterile neutrino, and  (c) assess the ability of these long baseline experiments to distinguish between the two scenarios.

The plan of the article is as follows: In Sec.~\ref{sec:framework}, we provide the framework used to study the role of eV-scale sterile neutrino in neutrino oscillations  and bring out the differences between the standard $(3+0)$ case and $(3+1)$ case  in terms of the $CP$ asymmetries (see Appendix~\ref{app_1} for the probability expressions for the $(3+1)$ case).
 In Sec.~\ref{sec:results},  we present numerical analysis and describe the consequences on $CP$ violation and discuss the role of appearance and disappearance channels. We also address the question of separating intrinsic versus extrinsic $CP$ effects both for  the $(3+0)$ case and $(3+1)$ case. In Sec.~\ref{sec:event}, we provide all the crucial details of the experiments (T2HK, DUNE and P2O) and fluxes  used in our numerical simulations. We finally address the key point of this article, i.e., if we can distinguish between the $(3+0)$ and $(3+1)$ case by defining a metric and by performing detailed numerical simulations at the level of event rates and $\chi^2$ (see Sec.~\ref{sec:chisq}). Further, we also deduce the expected precision on the active-sterile mixing angles at future long baseline experiments such as T2HK and DUNE (see Sec.~\ref{sec:precision}). We summarize and conclude in Sec.~\ref{sec:conclusion}. 
\section{Impact of eV-scale sterile neutrino at the probability level}
\label{sec:framework}
The Hamiltonian for the $(3+1)$ case in the flavor basis describing neutrino evolution governed by standard neutrino interactions is
\bea
\label{eq:ham}
{\mathcal{H}} \simeq \dfrac{1}{2E}~U \, \textrm{diag}\left(0,\sdm, \ldm, \lldm \right)U^\dagger + 
\textrm{diag}\left(
V_{CC}+V_{NC},V_{NC},V_{NC},0 \right)  \,.
\eea
where, $V_{CC} = \sqrt{2} G_{F} N_{e}$ is the effective charged-current (CC) potential (with $N_{e}$ being the electron number density) and $V_{NC} = -({1}/{\sqrt{2}}) G_{F} N_{n}$ is the neutral current (NC) potential (with $N_{n}$ being the neutron number density)~\footnote{It may be noted that only $\nu_{e}$ interact with the background $e$ (as there are no $\mu$ or $\tau$ in normal medium) via CC interactions whereas all active neutrino flavors interact with background matter ($e, p, n$) via NC interactions. For electrically neutral medium, the contributions due to $e$ and $p$ cancel out and thus only the $N_{n}$ appears in the NC contribution.}. $G_F=1.16\times 10^{-5}~{\textrm{GeV}}^{-2}$ is the Fermi constant. $U$ is the $4\times4$ unitary mixing matrix which can be parametrized in various ways~\cite{Kopp:2013vaa, Giunti:2019aiy, DUNE:2020fgq, Klop:2014ima}. In the present work, we  adopt the following parametrization~\cite{Klop:2014ima},
\bea
U &=&V(\theta_{34}, \delta_{34})O(\theta_{24})V(\theta_{14}, \delta_{14})O(\theta_{23})V(\theta_{13}, \delta_{13})O(\theta_{12})
\label{eq:umat}
\eea
where, $V(\theta_{ij},\delta_{ij})$ and $O(\theta_{ij})$ are complex and real rotation matrices  in the $i$-$j$ plane respectively (with $i , j = 1, 2, 3,4$). $U$ depends on six independent angles ($\theta_{12}, \theta_{13}, \theta_{23}, \theta_{14}, \theta_{24}, \theta_{34}$) and three independent phases ($\delta_{13}, \delta_{14}, \delta_{34}$)~\cite{ParticleDataGroup:2022pth}. The elements of $U$ are as follows:
 \bea
U &=& 
\begin{pmatrix}
c_{12} c_{13} c_{14} & c_{13} c_{14} s_{12} & c_{14}s_{13}\,{e}^{-{i} \delta_{14}}  &  s_{14}\,{e}^{-{i} \delta_{14}} \\
U_{\mu 1} & U_{\mu 2}  & U_{\mu 3}  &  c_{14} s_{24} \\
U_{\tau 1} & U_{\tau 2} & U_{\tau 2} & c_{14} c_{24} s_{34}\,{e}^{{-i} \delta_{34}} \\
U_{s 1} & U_{s 2} & U_{s 3} & c_{14} c_{24} c_{34} 
\label{eq:umat_n}
\end{pmatrix}
\eea
with,
\bea
U_{\mu 1} &=&(s_{24} s_{14} c_{13}{e}^{{i} \delta_{14}}-c_{24}s_{23}s_{13}{e}^{{i} \delta_{13}})c_{12} - c_{24} c_{23} s_{12} \,, \nonumber \\ 
U_{\mu 2} &=& c_{12} c_{23} c_{24}-s_{12}(s_{14} s_{24} c_{13}{e}^{-{i} \delta_{14}} +c_{24} s_{13} s_{23}{e}^{{i} \delta_{13}}) \,, \nonumber \\ 
U_{\mu 3} &=& c_{13} c_{24} s_{23}-c_{13} s_{14} s_{24} {e}^{{i}(\delta_{14}-\delta_{13})} \,, \nonumber \\
U_{\tau 1} &=& c_{12}\bigg[-s_{34} s_{14} c_{13}{e}^{{i}(\delta_{34}-\delta_{14})} -s_{13}{e}^{{i} \delta_{13}} (c_{34} c_{23} - s_{34} s_{24} s_{23}{e}^{{-i} \delta_{ 34}}) + s_{12}(s_{34} s_{24} c_{23}{e}^{{-i} \delta_{34}} +c _{34} s_{23})\bigg] \,, \nonumber \\ 
U_{\tau 2} &=& -c_{12}( c_{23} s_{24} s_{34}{e}^{{-i} \delta_{14}} + c_{34} s_{23}) -s_{12}\bigg[c_{24} s_{13} s_{14} s_{34}{e}^{{i}(\delta_{34}-\delta_{14})}-
s_{13} {e}^{{i} \delta_{13}} (c_{23} c_{34} - s_{23} s_{24} s_{34}{e}^{{-i} \delta_{34}})
\bigg] \,, \nonumber \\ 
U_{\tau 3} &=& c_{13}(c_{23} c_{34} -s_{23} s_{24} s_{34}{e}^{{i} \delta_{34}})- c_{24} s_{13} s_{14} s_{34} {e}^{{i} (\delta_{34}-\delta_{14} +\delta_{13})} \,,\nonumber \\  
U_{s1} &=& c_{12}\bigg[-c_{34} s_{14} c_{13} {e}^{{i} \delta_{14}}+s_{13}{e}^{{i} \delta_{13}} (s_{34} c_{23}+ c_{34} s_{24} s_{23} ) -s_{12}(-c_{34} s_{24} c_{23} {e}^{{i} \delta_{34}} +  s_{34} s_{23})\bigg] \,, \nonumber \\ 
U_{s2} &=& c_{12}(-c_{23}c_{34} s_{24} + s_{23} s_{34} {e}^{{i} \delta_{34}}) -s_{12}\bigg[c_{13} c_{24} c_{34} s_{14} {e}^{{i} \delta_{14}}- s _{13} {e}^{{i} \delta_{13}} (c_{23} s_{34} {e}^{{i} \delta_{34}} +  c_{34} s_{23} s_{24})\bigg] \,, \nonumber \\ 
U_{s3} &=& -c_{13}(c_{23} s_{34} {e}^{{i} \delta_{34}} + c_{34} s_{23} s_{24}) - c_{24} c_{34} s_{13} s_{14} {e}^{{i}(\delta_{14}-\delta_{13})} \nonumber.
\eea
where, $s_{ij} = \sin \theta_{ij}$ and $c_{ij} = \cos \theta_{ij}$. We can readily note that  when $\theta_{14}, \theta_{24}, \theta_{34}$ vanish, we recover the standard three flavor mixing matrix in the commonly adopted form~\cite{Pontecorvo:1957qd,Pontecorvo:1957cp,Gribov:1968kq,Maki:1962mu} (with $\delta_{13}$ identified as the Dirac-type $CP$ phase). In the limit when $\theta_{13}$, $\theta_{14}$, and $\theta_{24}$ are small, one gets 
$|U_{e3}|^2 \simeq s^2_{13}$, $|U_{e4}|^2 \simeq s^2_{14}$, $|U_{\mu4}|^2 \simeq s^2_{24}$, and $|U_{\tau4}|^2 \simeq s^2_{34}$ which allows for a clear physical interpretation of the new angles. 

Following the approach given in~\cite{Klop:2014ima}, we compute approximate analytical expressions of probability for the appearance channels $(\nu_{\mu}\to\nu_{e},\,\nu_{\mu}\to\nu_{\tau},\, \nu_{e}\to\nu_{\tau})$ and disappearance channels $(\nu_{e}\to\nu_{e},\,\nu_{\mu}\to\nu_{\mu},\, \nu_{\tau}\to\nu_{\tau})$  for $(3+1)$ case (see Appendix~\ref{app_1}).

\subsection{Review of the $(3+0)$ case in vacuum}
\label{sec:vacuum}

It is known that there is no $CP$ or $T$ violation in two flavor neutrino oscillations. However, in case of three flavor, there is a  complex phase in the $3\times 3$ mixing matrix and that may be responsible for observable $CP$ and $T$ violation effects in neutrino oscillations~\cite{Akhmedov:2004ve, KUO1987406}. The $3\times 3$ mixing matrix in the PMNS parameterization~\cite{Pontecorvo:1957qd,Pontecorvo:1957cp,Gribov:1968kq,Maki:1962mu} is given by
\bea
{{U}} &= 
\left( \begin{matrix} c_{12} c_{13} &
    s_{12} c_{13} & s_{13} {e}^{-{i} \delta_{13}} \\ -s_{12}
    c_{23} - c_{12} s_{13} s_{23} {e}^{{i} \delta_{13}} & c_{12}
    c_{23} - s_{12} s_{13} s_{23} {e}^{{i} \delta_{13}} & c_{13}
    s_{23} \\ s_{12} s_{23} - c_{12} s_{13} c_{23} {e}^{{i}
      \delta_{13}} & -c_{12} s_{23} - s_{12} s_{13} c_{23} {e}^{{i} \delta_{13}} & c_{13} c_{23} \end{matrix} \right) \,. 
   \label{eq:U}   
\eea
where $s_{ij} = \sin \theta_{ij}$, $c_{ij} = \cos \theta_{ij}$ and $\delta_{13}$ is the Dirac-type $CP$ violating  phase~\footnote{The two Majorana phases appear as a term proportional to Identity and therefore play no role in neutrino oscillations.}. Different parametrizations of the mixing matrix are possible and the impact of different parametrizations and interpretations for the $CP$ violating phase $\delta_{13}$ has been  discussed (see~\cite{Denton:2020igp}). 
Let us define the following probability differences corresponding to $CP$, $T$ and $CPT$ transformations:
\begin{subequations}
 \bea
 \label{eq:asymmdef}
     \Delta {P}_{\alpha\beta}^{CP}&=& {{P}_{\alpha \beta} - \bar {P}_{\alpha \beta}}\,, \label{CP}
     \quad  
    \\          
     \Delta {P}_{\alpha\beta}^{T}&=& {{P}_{\alpha \beta} -  {P}_{\beta \alpha}}\,, \label{T}
     \quad  
    \\            
      \Delta {P}_{\alpha\beta}^{CPT}&=& {{P}_{\alpha \beta} -  \bar {P}_{\beta \alpha}}\,. \label{CPT}
 \eea
 \end{subequations}
\noindent
where $P_{\alpha \beta}$ denotes transition probability for $\nu_\alpha \to \nu_\beta$ and $\bar P_{\alpha \beta}$ denotes  transition probability for $\bar\nu_\alpha \to \bar \nu_\beta$. {A schematic of these probability differences is depicted in Fig.~\ref{fig:def}.
\begin{figure}[t!]
\begin{center}
\includegraphics[width=4in]{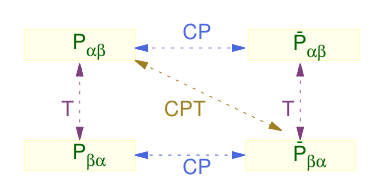}
\end{center}
\caption{Schematic of $CP$, $T$ and $CPT$ violation effects.}
\label{fig:def}
\end{figure}
Let us look at the consequences of $CPT$ conservation and three flavor unitarity in the context of three flavor neutrino oscillations in vacuum. From $CPT$ conservation, it follows that,
\bea
&&
\Delta {P}_{\alpha \beta}^{CP} = \Delta {P}_{\alpha \beta}^{T}\,;\,
\Delta {P}_{\alpha \alpha}^{CP}  = 0 = \Delta {P}_{\alpha \alpha}^{T}  \,. 
\label{eqn6}
\eea
Thus,  the $CP$ and $T$ violating probability differences are equal and there is no $CP$ and $T$ violation in survival (disappearance) channel. 
There are three $CP$ (and $T$) violating probability differences in three flavor case in vacuum which corresponds to the distinct appearance channels : $\Delta {P}_{\mu e}^{CP}$, $\Delta {P}_{ e\tau }^{CP}$ and $\Delta {P}_{\tau \mu}^{CP}$.  
Imposing three flavor unitarity $(\sum _{\beta}P_{\alpha\beta} = 1$ and $\sum_{\beta} \bar P_{\alpha\beta} = 1)$ leads to the following relation among the probability differences in the appearance channels,
\bea
\Delta {P}_{\mu e}^{CP} &=& \Delta {P}_{e \tau}^{CP} = -\Delta {P}_{\mu \tau }^{CP} 
\equiv {\Delta {P}}
\,,
\nonumber \\
\Delta {P}_{\mu e}^{T} &=& \Delta {P}_{e \tau}^{T} = -\Delta {P}_{\mu \tau }^{T}\,.
\equiv {\Delta {P}}
\label{eqn7}
\eea
This implies that the  $CP$  and $T$ violation probability differences are equal.
The $CP$ violating probability difference is  given by 
\begin{eqnarray} 
\Delta{P} &=& -16 \,{\mathcal{J}} \sin \Delta_{12} \sin \Delta_{23}  \sin \Delta_{31} \,,
\label{deltap}
\end{eqnarray}
where $\Delta_{ij} = {\Delta m^{2}_{ij} L}/{(4E)}$ (with $i , j = 1, 2, 3$). ${\mathcal{J}}$ is the Jarlskog factor given by  ${\mathcal{J}} = s_{13} c_{13}^{2} s_{12} c_{12} s_{23} c_{23} \sin \delta_{13}$ which is rephasing and parameterization independent measure of $CP$ violation~\cite{Jarlskog:1985ht}. 
Thus, Eq.~\eqref{deltap} leads to the following conditions for observability of non-trivial $CP$ and $T$ violating effects,
\begin{itemize}
\item  
$\delta_{13}$ $\neq 0^{\circ}$ or $180^{\circ}$
\item 
$\theta_{ij} \neq 0^{\circ}$ or $90^{\circ}$
\item 
$\Delta m^{2}_{ij} \neq 0$ 
\end{itemize}
Turning on matter effect destroys the nice feature of the equality of probability differences corresponding to $CP$ and $T$ violation so, Eq.~\eqref{eqn6} and Eq.~\eqref{eqn7} do not hold. It means that matter gives rise the effects which are different from the effects induced by $\delta_{13}$ (the so-called intrinsic effect). 
We will refer to these matter induced effects as fake/extrinsic effects.
When neutrinos propagate through matter, one must look at $CP$ (and $T$) violating effects in different appearance channels as well as  disappearance channels to conclude whether  $CP$ (or $T$) is violated or conserved in the leptonic sector.

As we are interested in the question of $CP$  violation effects and in particular the role played by matter, let us  take a look at the analytic expressions for $CP$ violating probability differences  both in the context of the $(3+0)$ case and $(3+1)$ case.

\subsection{$CP$ violating probability differences  in matter for the $(3+0)$ case}
\label{sec:matter(3+0)}
The approximate expression for probabilities in constant density matter (upto the second order in $\alpha = \Delta m^{2}_{21}/\Delta m^{2}_{31}$ and $\sin \theta_{13}$) are given in~\cite{Akhmedov:2004ny} (see also~\cite{Cervera:2000kp,Freund:2001pn,Minakata:2009sr, Agarwalla:2013tza, Minakata:2015gra, Masud:2015xva, Denton:2016wmg, Denton:2018hal, Ioannisian:2018qwl, Gronroos:2024jbs}). A comparison of accuracy and computational efficiency of various approximate expressions for probabilities is given in~\cite{Barenboim:2019pfp}. Following~\cite{Akhmedov:2004ny}, we can easily calculate~\footnote{Some of these expressions are available in literature (for instance, in~\cite{Freund:2001pn, Nunokawa:2007qh, Bass:2013vcg, Masud:2015xva}). We provide the expressions for all channels here for the sake of completeness.} the expressions for $\Delta P^{CP (3+0)}_{\alpha \beta}$ for the different appearance and disappearance channels. 

{For the {\bf{appearance channels}} we have,}
\begin{subequations}
\bea
\Delta P^{CP (3+0)}_{\mu e} &\simeq& 4 s_{13}^2  s_{23}^2  \Bigg[\frac{\sin^{2}(\hat A-1) \Delta}{(\hat A-1)^{2}} -  \frac{\sin^{2}(\hat A+1) \Delta}{(\hat A+1)^{2}} \Bigg] \nonumber \\ &-& 8\alpha\color{black}
\underbrace{\color{black}
  s_{13} s_{12} 
         s_{23} c_{12} 
         c_{23}\sin \delta_{13}
}_{{\mathcal{J}}/c^{2}_{13}}
\color{black} \frac{\sin \hat A \Delta}{\hat A} \Big[\cos \Delta \cot \delta_{13} \Theta_{-} + \sin \Delta \Theta_{+} \Big]  \,,
\label{eq:Delta P_em}\\ 
\Delta P^{CP (3+0)}_{\mu\tau} &\simeq& 2 s_{13}^2  \sin^{2} 2 \theta_{23}   \Bigg \{ \sin \Delta \cos \hat A \Delta \left( \frac{\sin(\hat A+1) \Delta}{(\hat A+1)^{2}} + \frac{\sin(\hat A-1) \Delta}{(\hat A-1)^{2}} \right)  \nonumber \\
&+& \frac{\hat A \Delta}{2} \Bigg( \frac{\sin 2\Delta}{\hat A+1} - \frac{\sin 2\Delta}{\hat A-1}\Bigg)  \Bigg \}
   + 8  \alpha \, \color{black}
\underbrace{\color{black}
  s_{13} s_{12} 
         s_{23} c_{12} 
         c_{23}\sin \delta_{13}
}_{{\mathcal{J}}/c^{2}_{13}}
\color{black} \sin\Delta \frac{\sin \hat A\Delta}{\hat A}
       \Theta_{+} \nonumber \\
          &-& \frac{1}{2 \hat A} \alpha^{2} \sin^{2} 2\theta_{12} \sin^{2} 2 \theta_{23}  \Bigg(\frac{1}{2\hat A} \sin 2\hat A \Delta \sin 2 \Delta - \Delta \sin 2\Delta \Bigg) \nonumber\\
         &-& 2\alpha \, s_{13} \sin 2\theta_{12} 
       \sin2\theta_{23}  \cos2\theta_{23}  \cos\delta_{13} 
       \sin\Delta \Bigg\{ \hat A \Bigg(\frac{\sin \Delta}{\hat A-1} - \frac{\sin \Delta}{\hat A+1} \Bigg) \nonumber \\
       &-& \frac{\sin \hat A\Delta}{\hat A} \Bigg(\frac{\cos(\hat A+1) \Delta}{\hat A+1} + \frac{\cos(\hat A-1) \Delta}{\hat A-1} \Bigg) \Bigg\}\, , 
\label{eq:Delta P_mt} \\ 
\Delta P^{CP (3+0)}_{e \tau} &\simeq& 4 s_{13}^2  s_{23}^2 \Bigg[\frac{\sin^{2}(\hat A-1) \Delta}{(\hat A-1)^{2}} -  \frac{\sin^{2}(\hat A+1) \Delta}{(\hat A+1)^{2}} \Bigg]\  \nonumber \\
         &+& 8  \alpha \color{black}
\underbrace{\color{black}
  s_{13} s_{12} 
         s_{23} c_{12} 
         c_{23}\sin \delta_{13}
}_{{\mathcal{J}}/c^{2}_{13}}
\color{black}  \frac{\sin \hat A \Delta}{\hat A}  \Big[\cos \Delta \cot \delta_{13} \Theta_{-} - \sin \Delta \Theta_{+} \Big]\,. 
\label{eq:Delta P_et}
\eea
\end{subequations}
{For the {\bf{disappearance channels}} we have,}
\begin{subequations}
\bea
   \Delta P^{CP(3+0)}_{ee} &\simeq& 4 s_{13}^2 \Bigg[\frac{\sin^{2}(\hat A+1) \Delta}{(\hat A+1)^{2}} -  \frac{\sin^{2}(\hat A-1) \Delta}{(\hat A-1)^{2}} \Bigg]  \,,
\label{eq:Delta P_ee} \\
\Delta P^{CP (3+0)}_{\mu\mu} &\simeq& \frac{1}{2\hat A} \alpha^2 \sin^2 2\theta_{12}  \sin^22\theta_{23} \Bigg[\frac{1}{2\hat A} \sin2 \hat A \Delta \sin 2 \Delta - \Delta \sin 2 \Delta \Bigg] \ \nonumber\\
   &+& 4 s_{13}^2 s_{23}^2 \Bigg[\frac{\sin^{2}(\hat A+1) \Delta}{(\hat A+1)^{2}} -  \frac{\sin^{2}(\hat A-1) \Delta}{(\hat A-1)^{2}} \Bigg]- 2 s_{13}^2 \sin^22\theta_{23}\Bigg\{\frac{\hat A \Delta}{2}
    \Bigg( \frac{\sin 2\Delta}{\hat A+1} - \frac{\sin 2\Delta}{\hat A-1}\Bigg)  \nonumber\\
    &+& \cos \Delta \cos \hat A \Delta \left(\frac{\sin(\hat A+1) \Delta}{(\hat A+1)^{2}} + \frac{\sin(\hat A-1) \Delta}{(\hat A-1)^{2}} \right) \Bigg \} \nonumber \\
   &+& 2 \alpha s_{13}  \sin 2\theta_{12}  \sin 2\theta_{23}
       \cos\delta_{13} \cos\Delta \frac{\sin \hat A\Delta}{\hat A} 
  \Theta_{-} \nonumber\\
   &+& 2\alpha s_{13} \sin 2\theta_{12}  
       \sin2\theta_{23} \cos2\theta_{23} \cos\delta_{13} 
   \sin\Delta \Bigg\{\hat A \Bigg(
 \frac{\sin \Delta}{\hat A-1} - \frac{\sin \Delta}{\hat A+1} \Bigg) \, \nonumber \\
 &-& \frac{\sin \hat A \Delta}{\hat A} \Bigg(\frac{\cos(\hat A+1) \Delta}{\hat A+1} + \frac{\cos(\hat A-1) \Delta}{\hat A-1} \Bigg) \Bigg\} \,,
\label{eq: Delta P_mm} \\
 \Delta P^{CP (3+0)}_{\tau \tau}&\simeq& 2 s_{13}^2 \, \sin^{2} 2 \theta_{23}   \Bigg\{ \sin \Delta \cos \hat A \Delta \left( \frac{\sin(\hat A+1) \Delta}{(\hat A+1)^{2}} + \frac{\sin(\hat A-1) \Delta}{(\hat A-1)^{2}} \right) \nonumber \\
 &+& \frac{\hat A \Delta}{2} \Bigg( \frac{\sin 2\Delta}{\hat A+1} - \frac{\sin 2\Delta}{\hat A-1}\Bigg)  \Bigg\} - \frac{1}{2\hat A} \alpha^{2} \sin^{2} 2\theta_{12} \sin^{2} 2 \theta_{23}  \Bigg(\frac{1}{2\hat A} \sin 2\hat A \Delta \sin 2 \Delta - \Delta \sin 2\Delta \Bigg) \nonumber\\
 &+& 8 \alpha \color{black}
\underbrace{\color{black}
  s_{13} s_{12} 
         s_{23} c_{12} 
         c_{23}\sin \delta_{13}
}_{{\mathcal{J}}/c^{2}_{13}}
\color{black} \sin\Delta \frac{\sin \hat A\Delta}{\hat A}
       \Theta_{+} 
       - 2\alpha s_{13} \sin 2\theta_{12}  
       \sin2\theta_{23} \cos2\theta_{23} \cos\delta_{13} 
       \sin\Delta \nonumber \\
       &&\Bigg\{\hat A \Bigg(\frac{\sin \Delta}{\hat A-1} - \frac{\sin \Delta}{\hat A+1} \Bigg)- \frac{\sin \hat A\Delta}{\hat A} \Bigg(\frac{\cos(\hat A+1) \Delta}{\hat A+1} + \frac{\cos(\hat A-1) \Delta}{\hat A-1} \Bigg) \Bigg\}\nonumber \\ 
       &+& 4 s_{13}^2 s_{23}^2 \Bigg[\frac{\sin^{2}(\hat A-1) \Delta}{(\hat A-1)^{2}} -  \frac{\sin^{2}(\hat A+1) \Delta}{(\hat A+1)^{2}} \Bigg] + 8 \alpha \color{black}
\underbrace{\color{black}
  s_{13} s_{12} 
         s_{23} c_{12} 
         c_{23}\sin \delta_{13}
}_{{\mathcal{J}}/c^{2}_{13}}
\color{black}\frac{\sin \hat A \Delta}{\hat A}  \nonumber \\ 
&& \Big[\cos \Delta \cot \delta_{13} \Theta_{-} - \sin \Delta \Theta_{+} \Big]\,.
        \label{eq:Delta P_tt} 
       \eea
\end{subequations}
Here
\bea
 \Delta & = & \dfrac{\Delta m^{2}_{31} L} {4E}\,,\,\alpha = \dfrac{\Delta m^{2}_{21}}{\Delta m^{2}_{31}}\,,\,\hat A = \dfrac{A}{{\Delta m^{2}_{31}}} \,,\, A = 2 \sqrt{2} G_F N_e E = 2  EV_{CC} \nonumber\\
  V_{CC} &=& \sqrt{2} G_F N_e \equiv7.64 \times10^{-14}  \dfrac{\rho}{[{\textrm{g/cc}}]}  Y_{e}~\textrm{eV},\nonumber \\
\Theta_{\pm} &=& \dfrac{\sin(\hat A+1) \Delta}{\hat A+1} \pm \dfrac{\sin(\hat A-1) \Delta}{\hat A-1} \,, \nonumber
\eea

Note that $N_e =N_{Avo} \rho Y_e $ with $N_{Avo} = 6.023 \times 10^{23} {\textrm{g}}^{-1} {\textrm{mole}}^{-1}$ being the Avogadro's number, $\rho$ being the matter density,  $Y_{e}$ being the number of electrons per nucleon, $Y_{e} \simeq 0.5$.
We can deduce the following :
\begin{itemize}
\item In case of vacuum, the $CP$ asymmetry was equal and independent of the particular appearance channel (see Eq.~\eqref{eqn7}). Incorporating matter effects destroy this nice feature of equality among the $CP$ asymmetries (see Eq.~\eqref{eq:Delta P_em}, Eq.~\eqref{eq:Delta P_mt} and Eq.~\eqref{eq:Delta P_et}) and we have
\bea
\Delta {P}_{\mu e}^{CP} &\neq& \Delta {P}_{e \tau}^{CP} \neq \Delta {P}_{\mu \tau }^{CP}\,.
\eea
\item 
The disappearance channels also depend on the $CP$ phase $\delta_{13}$. In general,  the $\delta_{13}$ term and matter term both appear in the disappearance probabilities. 
To the leading order, the approximate expression for $\Delta P^{CP (3+0)}_{ee}$  depends only on the matter potential and is independent of $\delta_{13}$ (see Eq.~\eqref{eq:Delta P_ee}) whereas $\Delta P^{CP (3+0)}_{\mu\mu}$ and $\Delta P^{CP (3+0)}_{\tau \tau}$  depend on $\delta_{13}$ as well as the matter potential (see Eq.~\eqref{eq: Delta P_mm} and Eq.~\eqref{eq:Delta P_tt}).
\end{itemize}

\subsection{$CP$ violating probability differences  in matter for the $(3+1)$ case}
\label{sec:matter(3+1)}
In $(3+1)$ case, there are distinct $CP$ violating probability differences both in appearance and disappearance channels (in contrast to the case of vacuum where Eq.~\eqref{eqn7} connects the probability differences of the appearance channels in vacuum)~\cite{Barger:1999hi, Kalliomaki:1999ii,Donini:1999he, Xing:2001bg, Guo:2001yt, Reyimuaji:2019wbn, Chatterjee:2022pqg}.   
The form of $\Delta P^{CP (3+1)}_{\alpha \beta}$ for the different channels (using the approximate probability expressions given in Appendix~\ref{app_1}) are given below:

 {For the {\bf{appearance channels}}, we have}
\begin{subequations}
\bea
\Delta P^{CP(3+1)}_{\mu e} &\simeq& 16{\hat{A}} s_{23}^2 s^2_{13}  \sin^2{\Delta}+  16 s_{13} s_{12} c_{12} s_{23} c_{23} (\alpha \Delta)\sin^2 \Delta \sin\delta_{13}\, \nonumber \\ 
&+& 4 s_{14} s_{24} s_{13} s_{23} \sin2\Delta \sin(\delta_{13}-\delta_{14})\cos\Delta \,, 
\label{eq:Delta P_mue_4nu} \\
\Delta P^{CP(3+1)}_{\mu \tau} &\simeq& \Delta P^{CP (3+0)}_{\mu \tau}+ \frac{2\hat{A}}{\hat{A}^2-1}\cos(\delta_{14}-\delta_{13})c_{23}\, \nonumber\\
&+& \cos2\theta_{23}\bigg[ \Omega_{-} \cos(\Delta+\delta_{13}+\omega_{-}) + \Omega_{+} \cos(\Delta-\delta_{13}+\omega_{+})\bigg]\, \nonumber \\
&-& \sin \theta_{13} \bigg[\Phi_{-} \sin(\Delta+\delta_{13}/2+ \omega_{-}) + \Phi_{+} \sin(\Delta-\delta_{13}/2+ \omega_{+})\bigg] \, \nonumber \\
&+& \frac{2\hat{A}}{\hat{A}^2-1} \cos(\delta_{13}-\delta_{14}-2(\hat{A}-1)\Delta)s_{13}s_{23}  
- c_{23}\bigg\{\Omega_{-} (\hat A -1) \cos\big(\delta_{34}+\omega_{-})-\delta_{13}/2 \big) \nonumber \\&+& \Omega _{+}(\hat A + 1) \cos\big(\delta_{34}-\omega_{+}) -\delta_{13}/2 \big) \bigg \}
- \Phi_{-} (\hat A - 1) \sin\big(\delta_{34}+ \omega_{-}) -\delta_{13}/2 \big) \nonumber \\ &+& \Phi_{+} (\hat A + 1)  \sin\big(\delta_{34}- \omega_{+}) + \delta_{13}/2 \big)
\,,\label{eq:Delta P_mutau_4nu}\\
\Delta P^{CP(3+1)}_{e \tau} &\simeq& c^2_{14}c^2_{24}\Delta P^{CP(3+0)}_{e \tau} + \frac{1}{2}c_{14}c_{24}s_{13}\sin 2\theta_{14}\sin 2\theta_{34}\bigg\{\frac{2\hat{A}}{\hat{A}-1}c_{23} \cos(\delta_{14}-2\delta_{13})  \, \nonumber \\ 
&+& 
 c_{23}\bigg(\Omega_{-}\cos(\delta_{14}-\delta_{13}/2+\omega_{-}) + \Omega_{+}\cos(\delta_{14}-\delta_{13}/2-\omega_{+})\bigg)  \nonumber \\
&& \bigg(\Phi_{+} \sin(\delta_{14}-\delta_{13}/2-\omega_{+})-
\Phi_{-} \sin(\delta_{14}-\delta_{13}/2+\omega_{-})
\bigg)
\bigg\}\,.
\label{eq:Delta P_etau_4nu}
\eea
\end{subequations}
Here
\bea
\Omega_{\pm} &=&  \frac{\cos(\Delta \pm \delta_{13}/2)}{\hat A \pm 1},\,\nonumber \Phi_{\pm} = \frac{\sin(\Delta \pm \delta_{13}/2)}{\hat A \pm 1},\,\,\omega_{\pm} = (1 \pm 2 \hat A ) \Delta\,. \nonumber
\eea
{For the {\bf{disappearance channels}}, we have}
\begin{subequations}
\bea
\Delta P^{CP(3+1)}_{e e} &\simeq& (1-2s^2_{14}) \Delta P^{CP(3+0)}_{ee}, 
\label{eq:Delta P_ee_4nu}\\
\Delta P^{CP(3+1)}_{\mu \mu} &\simeq& \Delta P^{CP(3+0)}_{\mu \mu}
 + \frac{1}{4}s_{13}s_{14}s_{24}c_{24}\bigg\{-s_{23}\bigg(\frac{\cos(\delta_{14}+(\hat{A}-1)\Delta)}{\hat{A}-1} \nonumber \\
 &+& \frac{\cos(\delta_{14}+(\hat{A}+1)\Delta)}{\hat{A}+1}\bigg)
 +\frac{2\hat{A}}{\hat{A}-1}s_{23}\cos(\delta_{14}+2\hat{A}\Delta)\, \nonumber \\
 &+& 2\sin(\delta_{13}+\delta_{14})\sin2\theta_{23}\bigg(\frac{\sin(2\Delta+\delta_{13})}{\hat{A}+1}-\frac{\sin(2\Delta-\delta_{13})}{\hat{A}-1}\bigg) + 2\sin3\theta_{23}\nonumber \\&&\bigg(\frac{\cos(2\Delta-\delta_{13})\cos\big(\delta_{14}-\delta_{13}/2-\omega_{-}\big)}{\hat{A}-1}+\frac{\cos(2\Delta+\delta_{13})\cos\big(\delta_{14}-\delta_{13}/2+\omega_{+} \big)}{\hat{A}+1}\bigg) \nonumber \\
&+& 2\sin 4\theta_{23}\cos(\delta_{14}+\delta_{13}) \bigg(\cos(2\Delta-\delta_{13})\Omega_{-} + \cos(2\Delta+\delta_{13})\Omega_{+} \bigg) 
 \bigg\} \nonumber \\
&+& \frac{1}{2} s_{14} s_{13} s_{23} s_{24} c_{24}\bigg\{\frac{\cos(\delta_{14}-2\Delta)}{\hat{A}-1}-\frac{\cos(\delta_{14}+2\Delta)}{\hat{A}+1}-\cos(\delta_{14}-\delta_{13})\bigg(\frac{2\hat{A}}{\hat{A}^2-1}\bigg) \nonumber \\
&+& 2\cos2\theta_{23}\bigg(\Omega_{-} \cos(\delta_{14}-\Delta-\delta_{13}/2)+ \Omega_{+} \cos(\delta_{14}+\Delta-\delta_{13}/2)\bigg)\bigg\} \,, 
\label{eq:Delta P_mumu_4nu}\\
\Delta P^{CP(3+1)}_{\tau \tau} &\simeq& c^4_{34}\Delta P^{CP(3+0)}_{\tau \tau}+ \frac{c^3_{34}s_{13}}{4(\hat{A}^2 - 1)}\bigg\{6\hat{A}\cos(\delta_{14}-3\delta_{13}) \nonumber \\
&-& 2(\hat{A}+1)\cos(\delta_{14}-2\delta_{13}-2\Delta)
 -2(\hat{A}-1)\cos(\delta_{14}-2\delta_{13}+2\Delta) \nonumber \\
 &+& 3 (\hat{A}+1)\cos(\delta_{14}-\delta_{13}-4\Delta) + 3 (\hat{A}-1)\cos(\delta_{14}-\delta_{13}+4\Delta) \nonumber \\ 
&+& 2\cos\theta_{23}\Big[3(\hat{A}-1)\cos(\delta_{14}+2(1-\hat{A})\Delta)\nonumber \\ &+& 3(\hat{A}+1)\cos(\delta_{14}-2(1+\hat{A})\Delta) - 2\hat{A}\cos(\delta_{14}-\delta_{13}-2\hat{A}\Delta)
\Big] \nonumber \\
&+& 8\cos2\theta_{23}\Big[\hat{A}\cos2\Delta\cos(\delta_{14}-\delta_{13})-\hat{A}\cos(\delta_{14}-3\delta_{13})+\sin2\Delta \sin(\delta_{14}-\delta_{13})
\Big] \nonumber \\
&+& 2\cos3\theta_{23}\Big[(\hat{A}-1)\cos(\delta_{14}+2(1-\hat{A})\Delta)+(\hat{A}+1)\cos(\delta_{14}-2(1+\hat{A})\Delta)\, \nonumber \\
&+& 2\hat{A}\cos(\delta_{14}-\delta_{13}-2\hat{A}\Delta)
\Big] + 4\cos4\theta_{23}\bigg(\hat{A}\cos(\delta_{14}-3\delta_{13})+2\cos2\Delta \cos(\delta_{14}-2\delta_{13})
\nonumber \\
&+&\cos(4\Delta)\cos(\delta_{14}-\delta_{13})+ 2 \sin2\Delta \sin(\delta_{14}-2\delta_{13})+\sin2\Delta \sin(\delta_{14}-\delta_{13})
\bigg)
\bigg\}\nonumber \\&+& {\cal O} (\epsilon^3)\,.
\label{eq:Delta P_tt_4nu}
\eea
\end{subequations}
We note the following for the $(3+1)$ case. 
\begin{itemize}
\item
The sterile sector introduces additional intrinsic sources of $CP$ violation in addition to matter effects.
\item From Eq.~\eqref{eq:Delta P_mue_4nu}, Eq.~\eqref{eq:Delta P_mutau_4nu} and Eq.~\eqref{eq:Delta P_etau_4nu}, we note that the leading order contribution to $CP$ violating probability difference in the appearance channels comes mainly from the $(3+0)$ contribution while the sub-leading terms depend on the active-sterile mixing. $\Delta P^{CP(3+0)}_{\mu \tau}$ and $\Delta P^{CP(3+0)}_{e \tau}$ are taken from Eq.~\eqref{eq:Delta P_mt} and Eq.~\eqref{eq:Delta P_et} respectively (such that we have terms upto ${\cal O} (\epsilon^3)$ in the final expression). 
\item The disappearance channels are sensitive to $CP$ violating effects (both intrinsic and extrinsic). 
From Eq.~\eqref{eq:Delta P_ee_4nu}, we note that $\Delta P^{CP(3+1)}_{ee}$ depends on  $\theta_{14}$ but not on $\delta_{13}$ or the sterile phases.  The $\Delta P^{CP(3+1)}_{\mu\mu}$  and $\Delta P^{CP(3+1)}_{\tau\tau}$ (Eq.~\eqref{eq:Delta P_mumu_4nu} and Eq.~\eqref{eq:Delta P_tt_4nu}) also depend on the respective $(3+0)$ terms at leading order while active-sterile contributions appear only at the sub-leading level.
\end{itemize}

In the next section, we numerically explore the behaviour of $CP$ violating probability differences and point out  the key distinguishing features in the $(3+1)$ case and $(3+0)$ case. The analytical expressions are useful in understanding the plots. 
\begin{figure}[t!]
 \begin{center}
\includegraphics[width=6.5in]{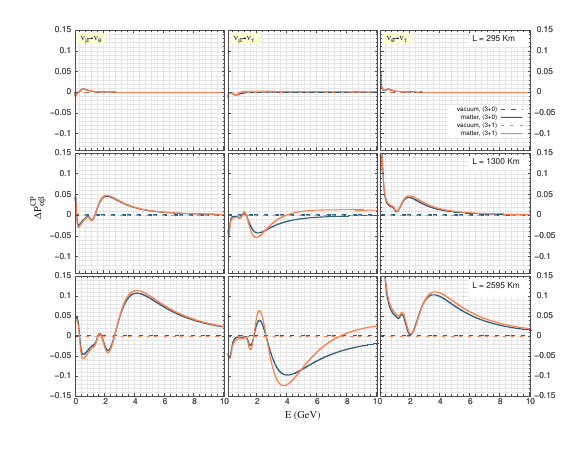}
\end{center}
\caption{$\Delta {P}^{CP}_{\alpha \beta}$ as a function of energy in $(3+0)$ case and $(3+1)$ case for $\delta_{13} = 0^{\circ}$. The three rows correspond to different baselines while the three columns correspond to the different appearance channels. Propagation in vacuum is depicted by dashed lines and propagation in matter is depicted by solid lines. We have chosen the sterile parameters as : $\Delta m^{2}_{41} = 1~\mbox{eV}^{2}$,  $\theta_{14} = 5.7^{\circ}$,  $\theta_{24} = 5^{\circ}$,  $\theta_{34} = 20^{\circ}$, $\delta_{14} = \delta_{34} = 0^{\circ}$.}
\label{fig1}
\end{figure}
\section{Numerical analysis at the probability level}
\label{sec:results}
We explore the behaviour of $CP$ 
violating differences at the level of probabilities as a function of energy, baseline and $\delta_{13}$. 
The simulations have been carried out using the software General Long Baseline Experiment Simulator (GLoBES)~\cite{Huber:2004ka,Huber:2007ji} which solves the full three flavor neutrino propagation equations numerically using the Preliminary Reference Earth Model (PREM)~\cite{Dziewonski:1981xy} density profile of the Earth.

\subsection{$CP$ violation and role of different oscillation channels}
\label{sec:cpviolation}
\begin{figure}[t!]
 \begin{center}
\includegraphics[width=6.5in]{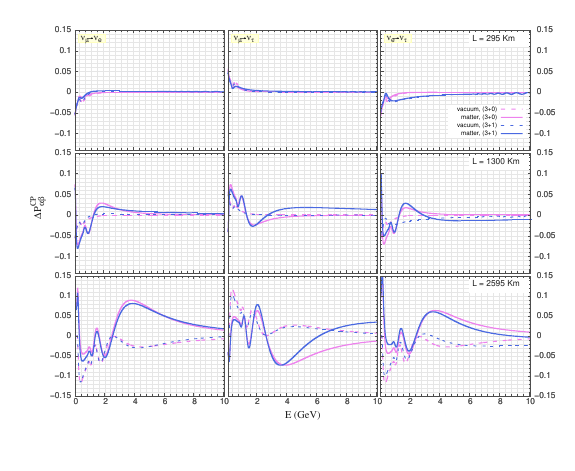}
\end{center}
\caption{Same as Fig.~\ref{fig1} but for $\delta_{13} = 90^{\circ}$.}
\label{fig2}
\end{figure}
\label{sec:3.1}
Let us first examine the $CP$ probability differences for appearance~(Fig.~\ref{fig1} and Fig.~\ref{fig2}) as well as disappearance channels ~(Fig.~\ref{CP_survival}). Different values of $\delta_{13}$ have been considered. 
For the appearance channels, $\Delta P^{CP}_{\alpha\beta}$ is plotted as a function of energy in Fig.~\ref{fig1} and Fig.~\ref{fig2} which correspond to $\delta_{13} =0^{\circ}$ and $\delta_{13} = 90^{\circ}$ respectively. The three rows correspond to the different baselines relevant for T2HK, DUNE and P2O. The three columns correspond to the different appearance channels. For the disappearance channels, there is no $CP$ asymmetry in vacuum and hence $\Delta P^{CP}_{\alpha\alpha}$ in matter is plotted as a function of energy in Fig.~\ref{CP_survival} for $\delta_{13} =0^{\circ}$ and $\delta_{13} = 90^{\circ}$. 
Below, we give how the particular channel is sensitive to the $CP$ violation effects.
\begin{figure}[t!]
 \begin{center}
\includegraphics[width=6.5in]{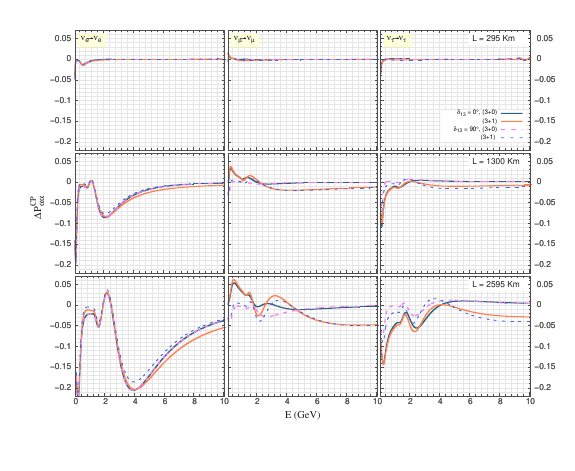}
\end{center}
\caption{$\Delta {P}^{CP}_{\alpha \alpha}$ in matter as a function of energy in $(3+0)$ case and $(3+1)$ case. The three rows correspond to different baselines while the three columns correspond to the different disappearance channels. $\delta_{13} = 90^{\circ}$  is depicted by dashed lines and $\delta_{13} = 0^{\circ}$ is depicted by solid lines. 
We have chosen the sterile parameters as : 
 $\Delta m^{2}_{41} = 1~\mbox{eV}^{2}$,  $\theta_{14} = 5.7^{\circ}$,  $\theta_{24} = 5^{\circ}$,  $\theta_{34} = 20^{\circ}$, $\delta_{14} = \delta_{34} = 0^{\circ}$.
}
\label{CP_survival}
\end{figure}

{For the {\bf{appearance channels}},}

\begin{enumerate}
\item[$\bullet$]
\textbf{$\nu_\mu \to \nu_e$ channel:} 

For the $(3+0)$ case in vacuum,  we expect $\Delta P^{CP}_{\mu e} = 0$ (dashed steel blue line in Fig.~\ref{fig1}) when $\delta_{13} = 0^{\circ}$.  For the $(3+1)$ case in vacuum,  again  $\Delta P^{CP}_{\mu e} = 0$ (dashed orange line in Fig.~\ref{fig1}) when $\delta_{13} = 0^{\circ}$. For the $(3+0)$ case in matter, $\Delta P^{CP}_{\mu e} \neq 0$ in general with some energy dependence and the extent of $CP$ asymmetry increases as the baseline increases. The $(3+1)$ case shows similar behaviour as the $(3+0)$ case in matter. 
When $\delta_{13} = 90^{\circ}$, we will have combination of intrinsic and extrinsic effects. Some differences begin to appear for $(3+0)$ and $(3+1)$ cases in vacuum. The $(3+1)$ case shows similar behaviour as the $(3+0)$ case in matter even for $\delta_{13} = 90^{\circ}$. 

\item[$\bullet$]
\textbf{$\nu_\mu \to \nu_\tau$ channel:}  

For the $(3+0)$ and $(3+1)$ case in vacuum,  we expect $\Delta P^{CP}_{\mu \tau} = 0$ (dashed steel blue line in Fig.~\ref{fig1}) when $\delta_{13} = 0^{\circ}$. 
In matter, when $\delta_{13} = 0^{\circ}$, there is discernible difference between $(3+0)$ and $(3+1)$ cases. 
For $\delta_{13} = 90^{\circ}$, we get $\Delta P^{CP}_{\mu \tau} \neq 0$ in vacuum with the sign being opposite to $\Delta P^{CP}_{\mu e}$ as expected (Eq.~\eqref{eqn7}). For $\delta_{13} = 90^{\circ}$, $\Delta P^{CP}_{\mu \tau} \neq 0$ in matter with the extent of asymmetry rising with baseline. We also note that some differences appear for the 
$(3+0)$ and $(3+1)$ cases in matter. 

\item[$\bullet$]
\textbf{$\nu_e \to \nu_\tau$ channel:} 

This channel exhibits similar behaviour as the $\nu_\mu \to \nu_e$ channel when $\delta_{13} = 0^{\circ}$ in both vacuum and matter. When $\delta_{13} = 90^{\circ}$, we can see the difference in the $(3+0)$ and $(3+1)$ cases depending upon the baseline and energy.
\end{enumerate}

{For the {\bf{disappearance channels}},}
\begin{enumerate}
\item[$\bullet$]
\textbf{$\nu_e \to \nu_e$ channel:} 

Fig.~\ref{CP_survival}  shows that  $\Delta P^{CP}_{ee}$ in matter is non-zero and independent of $\delta_{13}$ in the $(3+0)$ case  (as expected from Eq.~\eqref{eq:Delta P_ee}). However, at longer baselines $\Delta P^{CP}_{ee}$ has slight dependence on $\delta_{13}$ in the $(3+1)$ case (see Eq.~\eqref{eq:Delta P_ee_4nu}). The extent of the $CP$ asymmetry in this channel increases with baseline.

\item[$\bullet$]
\textbf{$\nu_\mu \to \nu_\mu$ channel:}
 
$\Delta P^{CP}_{\mu\mu}$ has a small dependence on $\delta_{13}$ at lower energies as can be seen from Fig.~\ref{CP_survival} (see Eq.~\eqref{eq: Delta P_mm} and Eq.~\eqref{eq:Delta P_mumu_4nu}). There is almost no $\delta_{13}$ dependence at shorter baselines.

\item[$\bullet$]
\textbf{$\nu_\tau \to \nu_\tau$ channel:} 

From Fig.~\ref{CP_survival}, we note that $\Delta P^{CP}_{\tau\tau}$ has a small dependence on $\delta_{13}$  as expected from Eq.~\eqref{eq:Delta P_tt}  and Eq.~\eqref{eq:Delta P_tt_4nu}.
\end{enumerate}

\subsection{Intrinsic versus extrinsic $CP$ violating effects in $(3+0)$ case and $(3+1)$ case at the probability level}
\label{sec:cpprob} 

We can define an observable quantity  by taking the difference of probabilities at $\delta_{13} = 0^{\circ}$ and $\delta_{13} = 90^{\circ}$ which allows us to distinguish between intrinsic and extrinsic $CP$ violating effects~\cite{Rout:2017udo,Nunokawa:2007qh} :
\begin{eqnarray}
\delta [\Delta P^{CP}_{\alpha\beta}]  &=& [\Delta P_{\alpha\beta}^{CP}] (\delta_{13} = 90^{\circ}) - [\Delta P_{
\alpha\beta}^{CP}] (\delta_{13}=0^{\circ})\,,\nonumber\\
\delta [\Delta P^{CP}_{\alpha\beta}]&=& [P_{\alpha\beta} - \bar P _{\alpha\beta}] (\delta_{13} = 90^{\circ}) - [P_{\alpha\beta} - \bar P _{\alpha\beta}] (\delta_{13}=0^{\circ})\,.
\label{eq:asysdef}
\end{eqnarray}
In vacuum, the second term on the RHS of Eq.~\eqref{eq:asysdef} vanishes and only the first term will contribute due to non-zero $\delta_{13}$ (Eq.~\eqref{eq:asysdef}  reduces to Eq.~\eqref{deltap} for this case). However in matter, there will be both intrinsic ($\delta_{13}$-induced) as well as extrinsic/fake contribution to  Eq.~\eqref{eq:asysdef}. Under certain conditions for the $\nu_\mu \to \nu_e$ channel~\cite{Marciano:2006uc}, the contribution due to matter effects becomes independent of the value of $\delta_{13}$ and it may be possible to decouple the intrinsic and extrinsic contributions near the peak of energy. However, in presence of eV-scale sterile neutrinos, such a decoupling may not work~\cite{Rout:2017udo}.

 \begin{table}[h]
\centering
\scalebox{0.9}{
\begin{tabular}{ c  c  c  c }
\hline
&&&\\
Parameter & True value & $3\,\sigma$ interval & $1\,\sigma$ uncertainty  \\
&&&\\
\hline
&&&\\
$\theta_{12}$ [$^\circ$]             & 34.3                    &  31.4 - 37.4   &  2.9\,\% \\
$\theta_{13}$ (NH) [$^\circ$]    & 8.53              &  8.16  -  8.94  &  1.5\,\% \\
$\theta_{13}$ (IH) [$^\circ$]    & 8.58              &  8.21  -  8.99   &  1.5\,\% \\
$\theta_{23}$ (NH) [$^\circ$]        & 49.3                     &  41.63  - 51.32    &  3.5\,\% \\
$\theta_{23}$ (IH) [$^\circ$]        & 49.5                     &  41.88  - 51.30    &  3.5\,\% \\
$\sdm$ [$\times 10^{-5} \text{eV}^2$]  & 7.5  &  [6.94 - 8.14]   &  2.7\,\% \\
$\ldm$ (NH) [$\times 10^{-3} \text{eV}^2$] & $+$ 2.56   &  [2.46 - 2.65] &  1.2\,\% \\
$\ldm$ (IH) [$\times 10^{-3} \text{eV}^2$] & $-$2.46  & $-$[2.37 - 2.55]  &  1.2\,\% \\
$\delta_{13}$ (NH) [$^\circ$] & $-$165.6   & [$-$180, 0]  $\cup$ [144, 180] &  $\ast$ \\
$\delta_{13}$ (IH) [$^\circ$]& $-$75.6   & [$-$154.8, $-$18]   & $\ast$  \\
$\Delta m^{2}_{41}$ [$\text{eV}^2$]  & 1.0 & $\ast$ & $\ast$ \\
$\theta_{14}$ [$^\circ$] & 5.7 & 0 - 18.4 & $\sigma(\sin^{2}\theta_{14}) = 5\,\%$ \\
$\theta_{24}$ [$^\circ$] & 5.0 & 0 - 6.05& $\sigma(\sin^{2}\theta_{24}) = 5\,\%$ \\
$\theta_{34}$ [$^\circ$] & 20.0 & 0 - 25.8 & $\sigma(\sin^{2}\theta_{34}) = 5\,\%$ \\
$\delta_{14}$ [$^\circ$]   & 0   & [$-$180, 180] &  $\ast$ \\
$\delta_{34}$ [$^\circ$]   & 0   & [$-$180, 180] &  $\ast$ \\
&&&\\
\hline
\end{tabular}}
\vspace {0.1in}
\caption{\label{parameters}
Oscillation parameters and corresponding uncertainties used in our study in the standard $(3+0)$ scenario~\cite{deSalas:2020pgw} and $(3+1)$ scenario~\cite{Dentler:2018sju}. If the $3\,\sigma$ upper and lower limits of a  parameter are $x_{u}$ and $x_{l}$ respectively, the $1\,\sigma$  uncertainty is $(x_{u}-x_{l})/3(x_{u}+x_{l})\%$~\cite{Abi:2020evt}. For the active-sterile mixing angles, a conservative $5\%$ uncertainty has been used on $\sin^{2}\theta_{i4}$ ($i = 1, 2, 3$).}
\end{table}
 \section{Experimental details, neutrino fluxes and results}
\label{sec:event}

The true values of oscillation parameters are given in Table~\ref{parameters}.  For the $(3+0)$ scenario, we take the best-fit values and the allowed $3\,\sigma$ range of oscillation parameters as given in~\cite{deSalas:2020pgw} (see also \cite{Esteban:2020cvm, Capozzi:2021fjo}).  These values have been obtained using  global analysis of neutrino oscillation data incorporating up-to-date experimental details.   For the $(3+1)$ scenario, we follow~\cite{Dentler:2018sju} where the authors have presented an updated global analysis of neutrino oscillation data within a $(3+1)$ sterile neutrino scheme and obtained updated constraints on the allowed strengths of mixing matrix elements $|U_{\alpha 4}|^{2}~(\alpha = e, \mu,\tau)$. In what follows, we provide the details of the long baseline experiments  (T2HK, DUNE and P2O)  considered in our work. 
\subsection{Details of  the long baseline experiments}
\label{sec:lbl}

\begin{table}[ht!]

\hspace*{1.4cm}
\begin{tabular}{p{2.5cm} p{10.5cm}}

 \hline
 \textbf{Experiment}   &  \hspace {2.5cm} \textbf{Details} \\
 \hline
          
   \textbf{T2HK}  & \textbf{$E_{p}$:} 30 GeV  \newline \textbf{{Baseline} (L):} 295 Km \newline\textbf{Runtime (year):} 2.5 $\nu$ + 7.5 $\bar \nu$  \newline  \textbf{Target mass (fiducial):} 187 kt \newline \textbf{Detector type:} WC \newline \textbf{Number of bins:} 95\newline \textbf{Bin width:} 0.05 GeV \newline $R_\mu=0.085/{\sqrt E}$, $R_e=0.085/{\sqrt E}$ \newline \textbf{Normalization error:} $\nu_{e}$ = 5\% (Sig), $\nu_{e}$ =  5\% (Bckg) \newline $\nu_{\mu}$ =  2.5\% (Sig), $\nu_{\mu}$ =  20\% (Bckg)\\
    
                 \hline

 \textbf{DUNE}     &  \textbf{$E_{p}$:} 120 GeV \newline \textbf{Baseline (L):} 1300 Km \newline \textbf{Runtime (year):}  6.5 $\nu$ + 6.5 $\bar \nu$ \newline \textbf{Target mass (fiducial):} 40 kt \newline \textbf{Detector type:} Liquid Argon Time Projection Chamber (LArTPC) \newline \textbf{Number of bins:} 64 \newline \textbf{Bin width:} 0.125 GeV \newline $R_\mu=0.20/{\sqrt E}$, $R_e=0.15/{\sqrt E}$ \newline \textbf{Normalization error:} $\nu_{e}$ = 2\% (Sig), $\nu_{e}$ =  5\% (Bckg) \newline $\nu_{\mu}$ =  5\% (Sig), $\nu_{\mu}$ =  5\% (Bckg)\\

  \hline

   \textbf{P2O}      &   \textbf{$E_{p}$:}  60 GeV \newline \textbf{Baseline (L):} 2595 Km \newline \textbf{Runtime (year):}  3 $\nu$ + 3 $\bar \nu$ \newline \textbf{Target mass (fiducial):} 4000 kt \newline \textbf{Detector type:} WC \newline \textbf{Number of bins:} 10 \newline \textbf{Bin width:} 1 GeV \newline $R_\mu=0.4/{\sqrt E}$, $R_e=0.4/{\sqrt E}$\newline \textbf{Normalization error:} $\nu_{e}$ = 5\% (Sig), $\nu_{e}$ =  10\% (Bckg) \newline $\nu_{\mu}$ =  5\% (Sig), $\nu_{\mu}$ =  10\% (Bckg)\\ 
 \hline
                \end{tabular}
\vspace {0.1in}
\caption{\label{sys} 
Details of the different experiments (T2HK, DUNE and P2O) considered in this work.}
\end{table}
\textbf{T2HK:} T2HK is planned neutrino oscillation experiment that consists of two detectors: a near detector (ND) at the Japan Proton Accelerator Research Complex (JPARC), located approximately $280$ m downstream from the neutrino production point, and a far detector (FD) based on water cherenkov (WC) technique in the Tochibora mines of Japan, $295$ Km away and about $1.7$ Km deep, located $8$ Km from Super-Kamiokande. 
 T2HK uses $30$ GeV proton beam with a beam power of $1.3$ MW, and will run both in $\nu$ and $\bar \nu$-mode. Combining  $\nu$ and $\bar \nu$ runs, the experiment will gather a total exposure of $\sim$ $2.7 \times 10^{22}$ protons on target (POT) which corresponds to  
\begin{eqnarray}
\dfrac{\textrm{POT/year}}{[2.7 \times 10^{22}]} \sim \dfrac{{\textrm{Proton beam power }}}{[1.3~{\textrm{MW}}]} \times \dfrac{T}{[10^{7} ~\textrm{sec}]} \times \dfrac{[30~\textrm{GeV}]}{E_p} 
\end{eqnarray}
We  divide the total POT in the ratio of $1:3$ for $\nu$ and $\bar \nu$-mode~\cite{Hyper-Kamiokande:2018ofw}. It corresponds to total runtime of $10$ year with the distribution of  $2.5$ year in the $\nu$-mode and  $7.5$ year in $\bar \nu$-mode, i.e., $2.5$($\nu$) + $7.5$($\bar \nu$). The total exposure comes around {\bf{2431 kt.MW.year}}. 

\textbf{DUNE:} 
DUNE is a upcoming long baseline experiment consisting of two detectors. The FD will be approximately $1300$ Km away and $1.5$ Km deep at the Sanford Underground Research Facility (SURF) in South Dakota and the ND system at baseline $0.570$ Km with target mass $0.067$ kt will be installed at the Fermi National Accelerator Laboratory (Fermilab), in Batavia, Illinois.  
The experimental design is capable of using a $120$ GeV proton beam with a beam power $1.2$ MW which corresponds to
\begin{eqnarray}
\dfrac{\textrm{POT/year}}{[1.1 \times 10^{21}]} \sim \dfrac{{\textrm{Proton beam power }}}{[1.2 ~{\textrm{MW}}]} \times \dfrac{T}{[10^{7} ~\textrm{sec}]} \times \dfrac{[120 ~\textrm{GeV}]}{E_p} 
\end{eqnarray} 
 We assume total run time of $13$ year, equally divided in $\nu$-mode and $\bar \nu$-mode, i.e., $6.5$($\nu$) + $6.5$($\bar \nu$)~\cite{DUNE:2021cuw}. The total exposure is around {\bf{624 kt.MW.year}}.

\textbf{P2O:} P2O is a long baseline experiment at a distance of $2595$ Km from the Protvino accelerator complex, which is located $100$ Km south of Moscow to the location of ORCA (Oscillation Research with Cosmics in the Abyss). This experimental setup employs a $90$ KW proton beam with energy between $50-70$ GeV which corresponds to 
\begin{eqnarray}
\dfrac{\textrm{POT/year}}{[0.8 \times 10^{20}]} \sim \dfrac{{\textrm{Proton beam power }}}{[90 ~{\textrm{KW}}]} \times \dfrac{T}{[10^{7} ~\textrm{sec}]} \times \dfrac{[60~\textrm{GeV}]}{E_p} 
\end{eqnarray}
We assume total run time of $6$ year with $3$ year in $\nu$-mode and $3$ year in $\bar\nu$-mode, i.e., $3$($\nu$) + $3$($\bar \nu$)~\cite{Akindinov:2019flp, KM3NET:2016zxf}. The total exposure is around {\bf{2160 kt.MW.year}}. 

The details of each experiment including the energy resolution and systematic effects are summarized in Table~\ref{sys}.

\subsection{Neutrino fluxes}
\label{sec:fluxes}

The flux of muon neutrinos (and anti-neutrinos) for the three different experiments is shown in Fig.~\ref{fig:flux}. The flux files have been taken from Ref.~\cite{Hyper-Kamiokande:2018ofw} for T2HK,  Ref.~\cite{DUNE:2021cuw} for DUNE and  Ref.~\cite{Akindinov:2019flp} for P2O respectively. 
We note (from left panel of Fig.~\ref{fig:flux}) that the spectrum for T2HK peaks at $E\simeq$ $0.6$ GeV which coincides with the first oscillation maximum of $\nu_\mu \to \nu_e$ probability for $ L =  295$ Km. The uncertainties in the flux arise due to the modeling of hadron production in graphite target and surrounding material~\cite{E899:1995bzq,T2K:2012bge}. 
For DUNE (see the middle panel of Fig.~\ref{fig:flux}), two different beam tunes have been used : (i) the standard low energy (LE) beam tune used in DUNE, Technical Design Report (TDR)~\cite{DUNE:2021cuw} and (ii) the medium energy (ME) beam tune optimized for $\nu_{\tau}$ appearance~\cite{dunefluxes}. The beams are generated from a G4LBNF simulation~\cite{Agostinelli:2002hh,Allison:2006ve} of the LBNF beam line using NuMI-style focusing. These two broad-band beam tunes  are consistent with what could be achieved by the LBNF facility. We have chosen a design for the ME beam that is nominally compatible with the space and infrastructure capabilities of the LBNF/DUNE beamline and that is based on a real working beamline design currently deployed in NuMI/NOvA. It may be noted that the LE flux peaks around  $E\simeq$ $2.5$ GeV (which coincides with the first oscillation maximum of $\nu_\mu \to \nu_e$ probability) for $ L =  1300$ Km while the ME flux peak is slightly shifted to around $5$ GeV.
The flux corresponding to P2O is shown in the right panel of Fig.~\ref{fig:flux} assuming zero-off axis angle. The factors like scattering, absorption and energy loss of hadron have been considered as per the beamline design~\cite{Abramov:2001nr}. The flux is peaked at  $E \simeq$ $5.1$ GeV for $ L =  2595$ Km which coincides with the first oscillation maximum of $\nu_\mu \to \nu_e$ probability.

  \begin{figure*}[t!]
\includegraphics[width=2.1in]{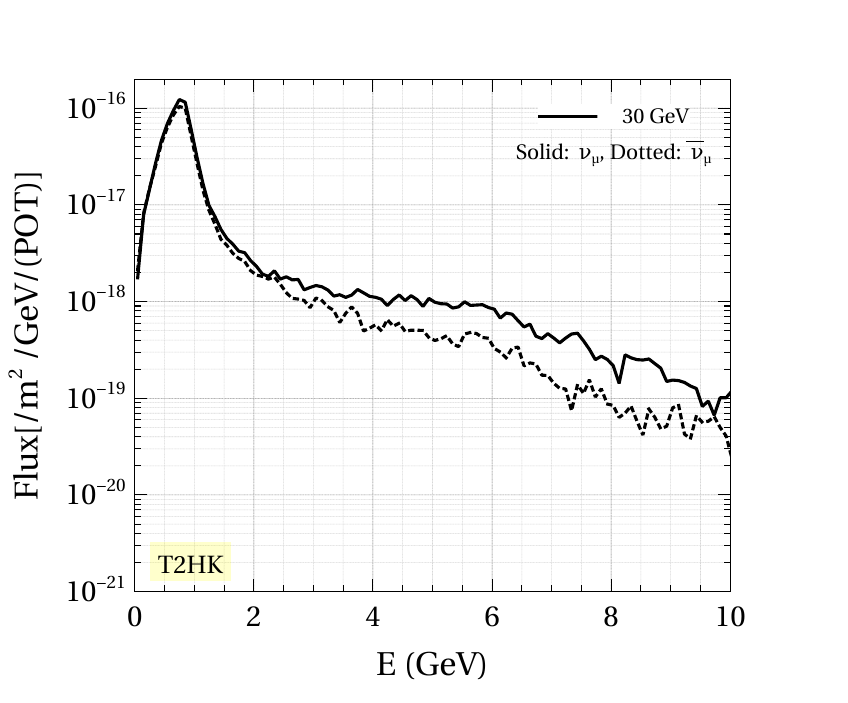}
\includegraphics[width=2.1in]{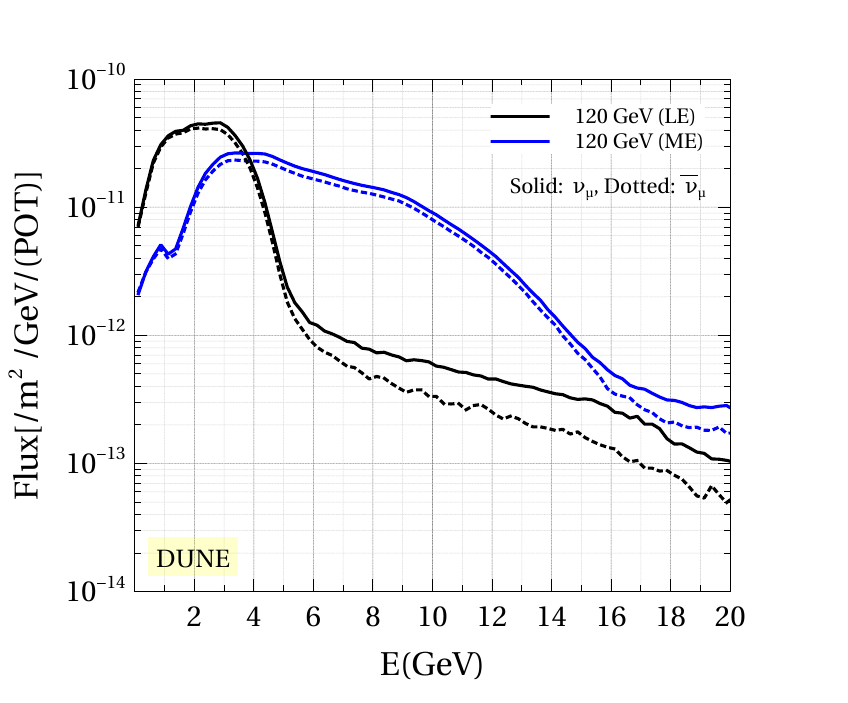}
\includegraphics[width=2.1in]{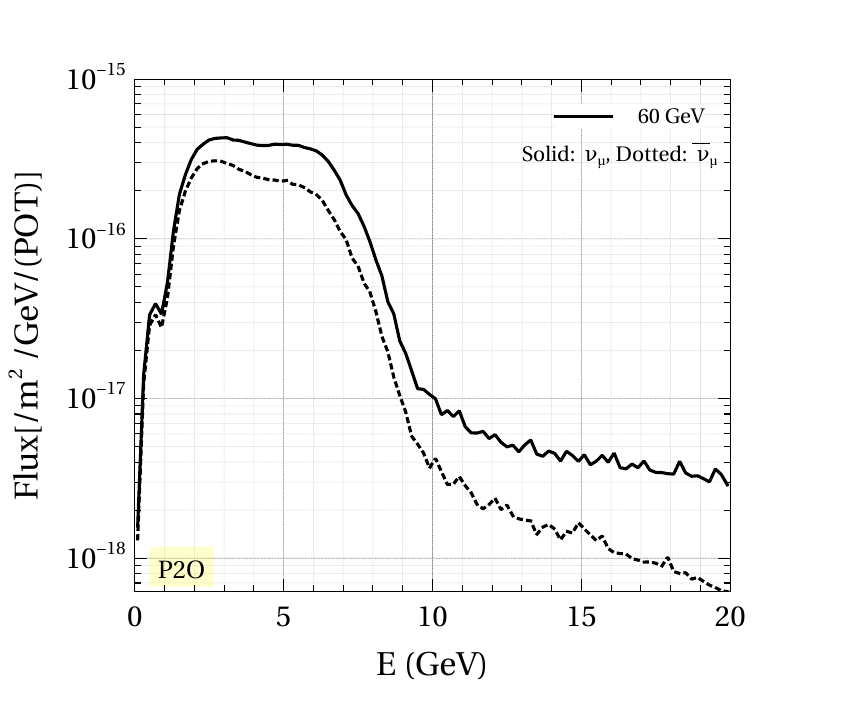}
\caption{Neutrino fluxes corresponding to the three experiments (T2HK, DUNE and P2O) considered in the present study. The flux plots have been adapted from~\cite{Hyper-Kamiokande:2018ofw,DUNE:2021cuw,Akindinov:2019flp}. For more details, see Ref.~\cite{Hyper-Kamiokande:2018ofw} for T2HK,  Ref.~\cite{DUNE:2021cuw} for DUNE and  Ref.~\cite{Akindinov:2019flp} for P2O. }
\label{fig:flux}
\end{figure*} 

\subsection{Intrinsic versus extrinsic $CP$ violating effects in $(3+0)$ case and $(3+1)$ case at the level of event rates}
\label{sec:cpevent}

As mentioned in Sec.~\ref{sec:cpprob}, distinguishing between  intrinsic and extrinsic $CP$ violating effects is one of the important goals of current and future long baseline experiments and  Eq.~\eqref{eq:asysdef} was proposed as a useful observable in this regard. Following Eq.~\eqref{eq:asysdef}, we can  define a quantity,
\bea
\dn  = [\Delta N_{\alpha\beta}^{CP}] (\delta_{13} = 90^{\circ}) - [\Delta N_{
\alpha\beta}^{CP}] (\delta_{13}=0^\circ)\,,
\label{event_def}
\eea
with $[\Delta N_{\alpha \beta}^{CP}]$ given by
\bea
[\Delta N_{\alpha \beta}^{CP}] \equiv 
[N_{\alpha \beta}
- \bar N_{\alpha \beta}]\,.
\label{quantity}
\eea
 $N_{\alpha \beta}$ can be expressed as, 
\bea
N_{\alpha \beta}~(L) &=& N_{target} \times  \int  
 \Phi_{\nu_\alpha} (E,L)  \times P_{\alpha \beta} (E,L) \times \sigma_{\nu_\beta} (E)~dE \, ,
\label{eventeqn}
\eea
 where
  $N_{target}$ is the number of target nucleons per kiloton of detector fiducial volume, 
  $P_{\alpha \beta}  (E,L)$ is the oscillation probability in matter. 
  $N^{T2HK}_{target}=6.252 \times 10^{33}~N/\textrm{kt}$, $N^{DUNE}_{target}$ = $6.022 \times 10^{32}~N/\textrm{kt}$ and $N^{P2O}_{target}=1.337 \times 10^{35}~N/\textrm{kt}$ where, $N$ stands for the number of nucleons. $\Phi_{\nu_\alpha} (E,L)$ is the flux of $\nu_\alpha$,  $\sigma _{\nu_\beta} (E)$ is the charged current (CC) cross section of $\nu_\beta$. For anti-neutrinos, $N_{\alpha \beta} \to \bar N_{\alpha \beta}$, $\nu_\alpha \to \bar \nu_\alpha$, $\nu_\beta \to \bar \nu_\beta$~\cite{Bass:2013vcg, Masud:2016nuj}.
Here  $\sigma _{\nu_e}(E)$ is given by, 
\bea
\label{eq:cross_sec}
 \sigma_{\nu_e}(E) &=& 0.67 \times 10^{-42} (m^2/{\rm{GeV}}/N) \times E\,, \quad {\rm {for}} \quad E > 0.5\,{\rm{GeV}}
\eea
It may be noted that $\sigma_{\nu_{\mu}} \simeq \sigma_{\nu_{e}}$ for the energy range considered. For anti-neutrinos, the cross-section is roughly a factor of two smaller~\cite{Hyper-Kamiokande:2018ofw, DUNE:2021cuw, Akindinov:2019flp}.
 \begin{figure}[t!]
 \begin{center}
\includegraphics[width=6.5in]{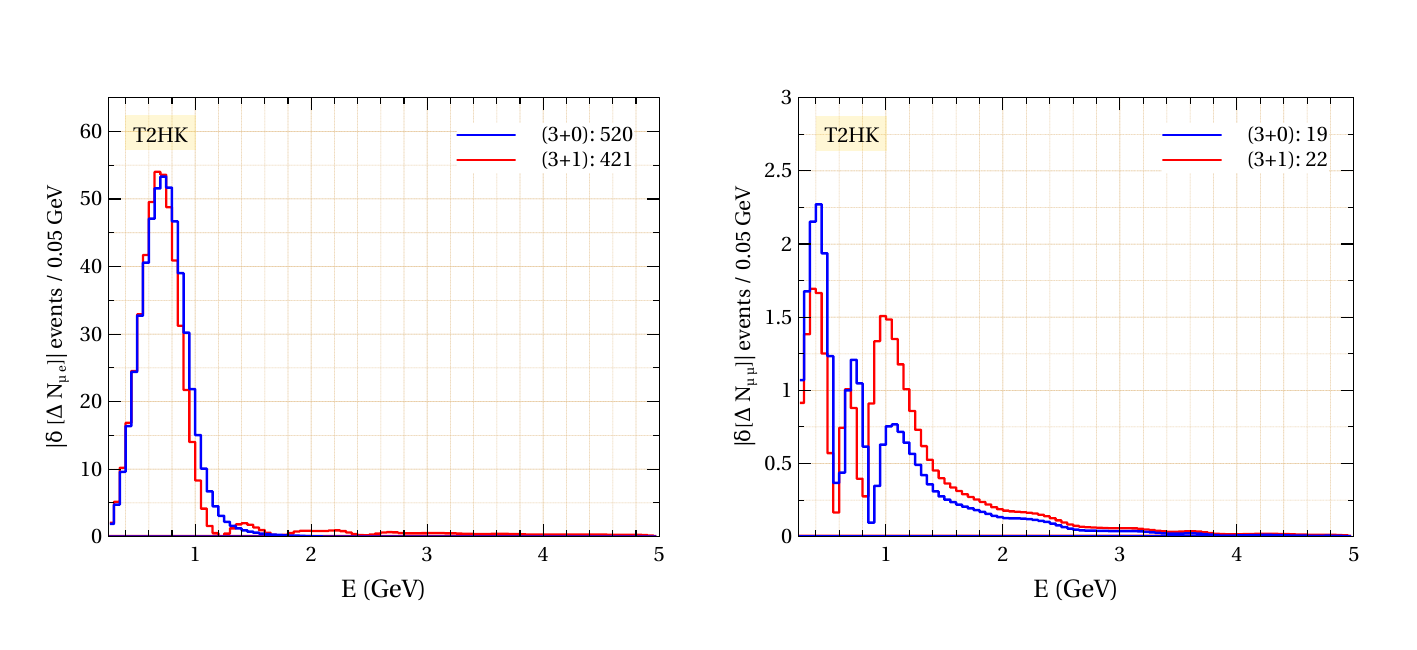}
\end{center}
\caption{$\delta [\Delta N_{\mu e}]$ and $\delta [\Delta N_{\mu \mu}]$ plotted as function of energy for T2HK. Both $(3+0)$ and $(3+1)$ cases are shown in each plot and the total event rate is also indicated for each case. }
\label{fig:event_def1}
\end{figure}
\begin{table}[ht!]
\hspace*{-0.7cm}
{\small{
\begin{tabular}{|p{3.8cm}|p{1.2cm}|p{1.2cm}|p{1.2cm}|p{1.2cm}|p{1.2cm}|p{1.2cm}|p{1.2cm}|p{1.2cm}|}
\hline T2HK
&\multicolumn{4}{c|}{(3+0) case} & \multicolumn{4}{c|}{(3+1) case} \\
\hline
&&&&&&&&\\
 & $N _{\mu e}$ &  $\bar{N} _{\mu e}$& $N _{\mu \mu}$ &  $\bar{N} _{\mu \mu}$&  $N _{\mu e}$ &  $\bar{N} _{\mu e}$&$N _{\mu \mu}$ &  $\bar{N} _{\mu \mu}$   \\
\hline
Sig+Bckg ($\delta_{13} = 0^{\circ}$)     &\textbf{2574}   &4222   &13581  &17349       &\textbf{2616}   &4306   &13332  &17188    \\
Bckg ($\delta_{13} = 0^{\circ}$)      &1355   &2171  &254   &238   &1336   &2135  &254    &238        \\
Sig+Bckg ($\delta_{13} = 90^{\circ}$)  &2297   &\textbf{4465}   &\textbf{13673}  &\textbf{17460}   &2293   &\textbf{4404}   &\textbf{13456}  &\textbf{17334}      \\
Bckg ($\delta_{13} = 90^{\circ}$)    &1356  &2171   &254   &238       &1336  &2136   &254    &238     
\\
 \hline
 \end{tabular}}}
 \vspace {0.1in}
\caption{Total  neutrino ($N_{\alpha \beta}$) and anti-neutrino ($\bar N _{\alpha \beta}$) event rate at T2HK corresponding to $\nu_{e}$ appearance and $\nu_{\mu}$ disappearance channels  for the $(3+0)$ and $(3+1)$ case. }
\label{table1_t2hk}
\end{table}

For computing the events for the experiments mentioned in Sec.~\ref{sec:lbl}, we use the software GLoBES~\cite{Huber:2004ka,Huber:2007ji} along with appropriate fluxes (see Sec.~\ref{sec:fluxes}) and cross-section as mentioned above (see Eq.~\eqref{eq:cross_sec}).  We consider both appearance ($\nu_\mu \to \nu_e$, $\bar \nu_\mu \to \bar \nu_e$) and disappearance channels ($\nu_\mu \to \nu_\mu$, $\bar \nu_\mu \to \bar \nu_\mu$). 
The backgrounds include $\nu_{\mu}\to\nu_{e}/\nu_{\tau}$ and $\nu_{\mu}\to\nu_{\mu}$ CC in $\nu$ and $\bar{\nu}$-mode as well as neutral current (NC), $\nu_\mu / \nu_e \to X$ (NC) for appearance and disappearance channels (see details in~\cite{Hyper-Kamiokande:2018ofw, DUNE:2021cuw, Akindinov:2019flp}). In generating event rates, the systematic uncertainties for each flavor $(\nu_{e}, \nu_{\mu}, \nu_{\tau})$ has been incorporated (see Table~\ref{sys}). The signal as well as background events are tabulated in Table~\ref{table1_t2hk} (T2HK), Table~\ref{table2_LE} (DUNE, LE), Table~\ref{table3_ME} (DUNE, ME) and Table~\ref{table4_p2o} (P2O) (with the largest number of events in a given channel depicted in boldface) for different experimental configurations listed in Table~\ref{sys}. The event spectra (corresponding to Eq.~\eqref{event_def}) are depicted in Fig.~\ref{fig:event_def1} for T2HK, Fig.~\ref{fig:event_def2} and Fig.~\ref{fig:event_def3} for DUNE and Fig.~\ref{fig:event_def4} for P2O. 

In Fig.~\ref{fig:event_def1}, $\delta [\Delta N_{\mu e}]$ and $\delta [\Delta N_{\mu \mu}]$  are plotted as a function of energy for T2HK. For $\nu_{\mu} \to \nu_{e}$ channel, one may be able to distinguish the intrinsic and extrinsic contributions in the lower energy range (below $1.5$ GeV). For $\nu_{\mu} \to \nu_{\mu}$ channel, $\delta [\Delta N_{\mu \mu}]$ is in general much smaller for all energies considered and it is very hard to distinguish between the intrinsic and extrinsic contributions. The signal and background events for T2HK are tabulated in Table~\ref{table1_t2hk} for $(3+0)$ and $(3+1)$ case respectively. 

For DUNE, we use two beam tunes (LE and ME)~\cite{DUNE:2021cuw, dunefluxes}. In Fig.~\ref{fig:event_def2}, we show $\delta [\Delta N_{\mu e}]$
plotted as a function of energy. The left panel is for $(3+0)$ case and right panel is for $(3+1)$ case. It is clear that the spectrum of $\delta [\Delta N_{\mu e}]$ depends on the flux used both for  $(3+0)$ and $(3+1)$ case. In Fig.~\ref{fig:event_def3}, we plot $\delta [\Delta N_{\mu \mu}]$ 
as a function of energy. The signal and background events for DUNE are tabulated in Table~\ref{table2_LE} and Table~\ref{table3_ME} for the $(3+0)$ and $(3+1)$ cases. 
 
\begin{figure}[t!]
\begin{center}
\includegraphics[width=6.5in]{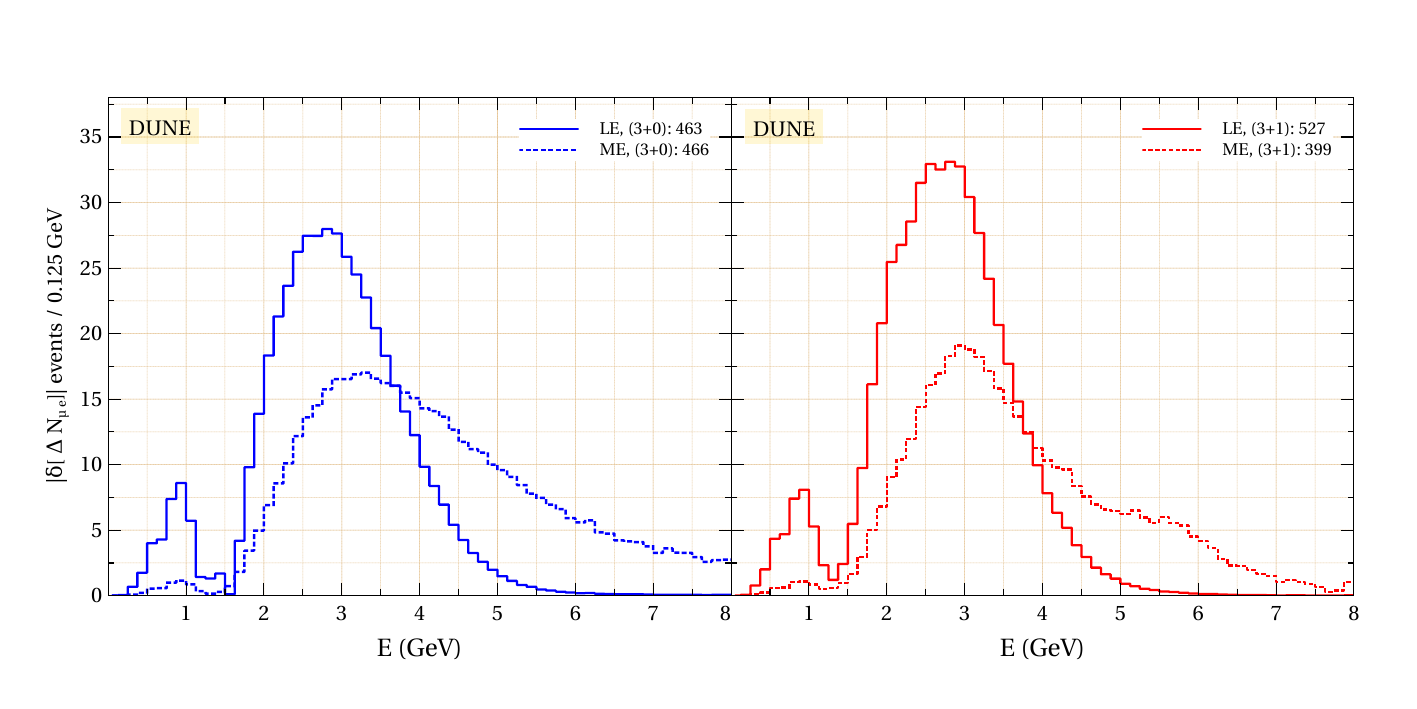}
\end{center}
\caption{$\delta [\Delta N_{\mu e}]$  plotted as function of energy for DUNE  (both LE and ME) and for  $(3+0)$  and $(3+1)$ cases. The total event rate is also indicated for each case.}
\label{fig:event_def2}
\end{figure}
\begin{figure}[t!]
 \begin{center}
\includegraphics[width=6.5in]{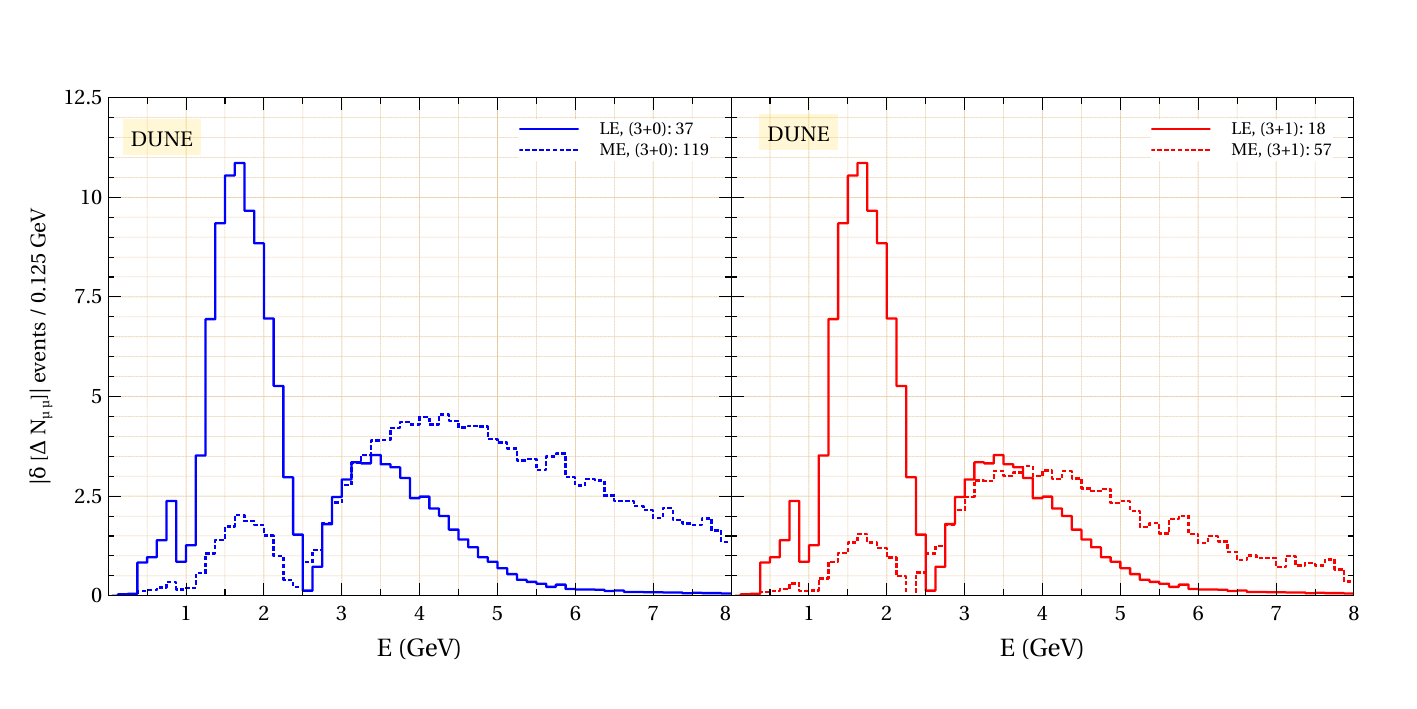}
 \end{center}
\caption{$\delta [\Delta N_{\mu \mu}]$  plotted as function of energy for DUNE  (both LE and ME). Both $(3+0)$ and $(3+1)$ cases are shown in each plot and the total event rate is also indicated for each case.}
\label{fig:event_def3}
\end{figure}
%
\begin{figure}[t!]
\begin{center}
\includegraphics[width=6.5in]{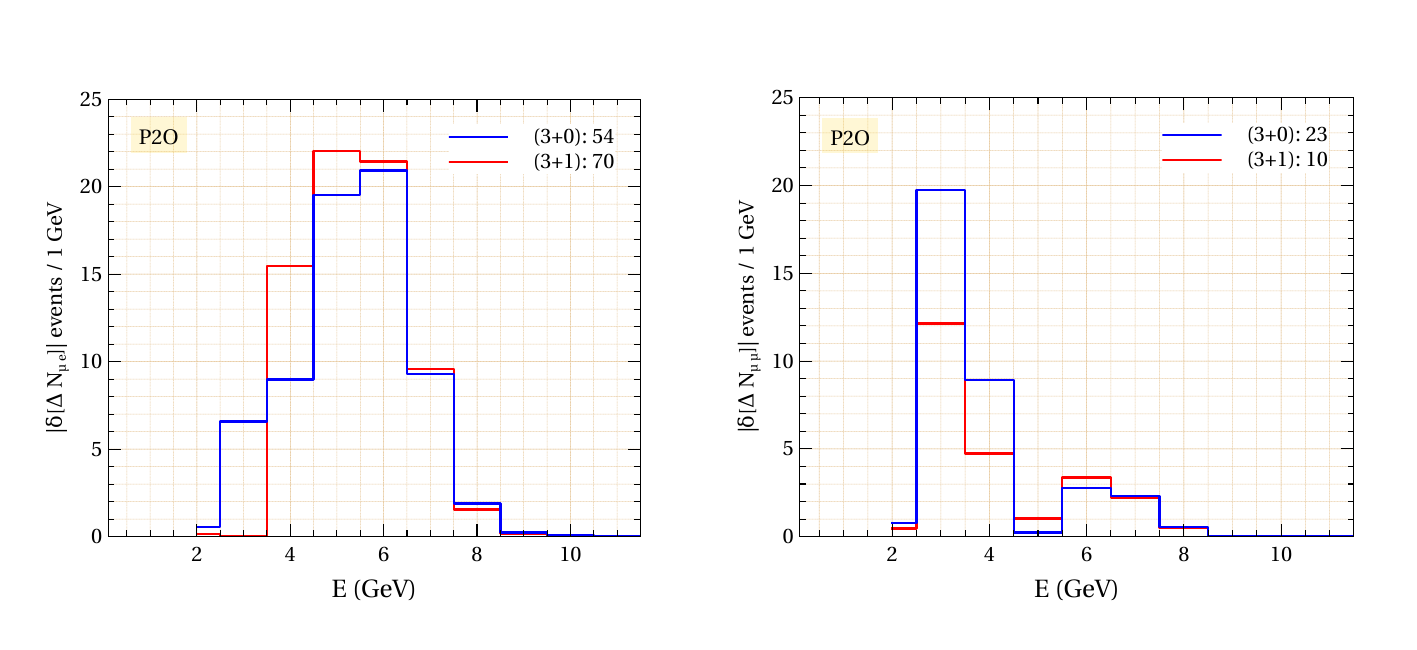}
\end{center}
\caption{$\delta [\Delta N_{\mu e}]$ and $\delta [\Delta N_{\mu \mu}]$  plotted as function of energy for P2O. Both $(3+0)$ and $(3+1)$ cases are shown in each plot and the total event rate is also indicated for each case.}
\label{fig:event_def4}
\end{figure}
\begin{table}[ht!]
\hspace*{-0.7cm}
{\small{
\begin{tabular}{|p{3.8cm}|p{1.2cm}|p{1.2cm}|p{1.2cm}|p{1.2cm}|p{1.2cm}|p{1.2cm}|p{1.2cm}|p{1.2cm}|}
\hline DUNE (LE)
&\multicolumn{4}{c|}{(3+0) case} & \multicolumn{4}{c|}{(3+1) case} \\
\hline
&&&&&&&&\\
 & $N _{\mu e}$ &  $\bar{N} _{\mu e}$& $N _{\mu \mu}$ &  $\bar{N} _{\mu \mu}$&  $N _{\mu e}$ &  $\bar{N} _{\mu e}$&$N _{\mu \mu}$ &  $\bar{N} _{\mu \mu}$   \\
\hline
Sig+Bckg($\delta_{13} = 0^{\circ}$)     &\textbf{2683}   &884   &\textbf{13069}  &6570       &\textbf{2791}   &927   &12664  &6496    \\
Bckg ($\delta_{13} = 0^{\circ}$)      &550   &314  &387    &223   &540   &309  &387    &223        \\
Sig+Bckg ($\delta_{13} = 90^{\circ}$)  &2284   &\textbf{947}   &13053  &\textbf{6591}   &2314   &\textbf{978}   &\textbf{12676}  &\textbf{6527}      \\
Bckg ($\delta_{13} = 90^{\circ}$)    &550  &314   &387    &223      &541  &309   &387    &223     
\\
 \hline
 \end{tabular}}}
 \vspace {0.1in}
\caption{Total  neutrino ($N_{\alpha \beta}$) and anti-neutrino ($\bar N _{\alpha \beta}$) event rate at DUNE (LE beam) corresponding to $\nu_{e}$ appearance and $\nu_{\mu}$ disappearance channels  for the $(3+0)$ and $(3+1)$ case.}
\label{table2_LE}
\end{table}
\begin{table}[ht!]
\hspace*{-0.7cm}
{\small{
\begin{tabular}{|p{3.8cm}|p{1.2cm}|p{1.2cm}|p{1.2cm}|p{1.2cm}|p{1.2cm}|p{1.2cm}|p{1.2cm}|p{1.2cm}|}
\hline DUNE (ME)
&\multicolumn{4}{c|}{(3+0) case} & \multicolumn{4}{c|}{(3+1) case} \\
\hline
&&&&&&&&\\
 & $N _{\mu e}$ &  $\bar{N} _{\mu e}$& $N _{\mu \mu}$ &  $\bar{N} _{\mu \mu}$&  $N _{\mu e}$ &  $\bar{N} _{\mu e}$&$N _{\mu \mu}$ &  $\bar{N} _{\mu \mu}$   \\
\hline
Sig+Bckg ($\delta_{13} = 0^{\circ}$)     &\textbf{3099}   &1062   &40428  &18174       &\textbf{3223}   &1114   &39206  &18124   \\
Bckg ($\delta_{13} = 0^{\circ}$)      &939   &435  &1314    &629  &930   &432  &1314    &630        \\
Sig+Bckg ($\delta_{13} = 90^{\circ}$)  &2662   &\textbf{1092}   &\textbf{40619}  &\textbf{18246}   & {2805}  &\textbf{1097}   &\textbf{39373}  &\textbf{18219}      \\
Bckg ($\delta_{13} = 90^{\circ}$)    &939 &435   &1314    &630       &931  &432   &1314    &630     
\\
 \hline
 \end{tabular}}}
 \vspace {0.1in}
\caption{Total  neutrino ($N_{\alpha \beta}$) and anti-neutrino ($\bar N _{\alpha \beta}$) event rate at DUNE (ME beam) corresponding to $\nu_{e}$ appearance and $\nu_{\mu}$ disappearance channels  for the $(3+0)$ and $(3+1)$ case.}
\label{table3_ME}
\end{table}
In Fig.~\ref{fig:event_def4}, $\delta [\Delta N_{\mu e}]$ and $\delta [\Delta N_{\mu \mu}]$  are plotted as a function of energy for P2O. For $\nu_{\mu} \to \nu_{e}$ and $\nu_{\mu} \to \nu_{\mu}$ channels, one may be able to distinguish between intrinsic and extrinsic effects at certain values of energy (around $2$ - $8$ GeV).
The signal and background events for P2O are tabulated in Table~\ref{table4_p2o} for $(3+0)$ and $(3+1)$ case respectively.
\begin{table}[ht!]
\hspace*{-0.7cm}
{\small{
\begin{tabular}{|p{3.8cm}|p{1.2cm}|p{1.2cm}|p{1.2cm}|p{1.2cm}|p{1.2cm}|p{1.2cm}|p{1.2cm}|p{1.2cm}|}
\hline P2O
&\multicolumn{4}{c|}{(3+0) case} & \multicolumn{4}{c|}{(3+1) case} \\
\hline
&&&&&&&&\\
 & $N _{\mu e}$ &  $\bar{N} _{\mu e}$& $N _{\mu \mu}$ &  $\bar{N} _{\mu \mu}$&  $N _{\mu e}$ &  $\bar{N} _{\mu e}$&$N _{\mu \mu}$ &  $\bar{N} _{\mu \mu}$   \\
\hline
Sig+Bckg ($\delta_{13} = 0^{\circ}$)     &\textbf{870}   &150   &1395  &453       &\textbf{871}   &148   &1353  &445   \\
Bckg ($\delta_{13} = 0^{\circ}$)      &417   &129  &296    &89   &400   &124  &281    &84        \\
Sig+Bckg ($\delta_{13} = 90^{\circ}$)  &820   &\textbf{154}   &\textbf{1374}  &\textbf{454}   & {808}  &\textbf{155}   &\textbf{1348}  &\textbf{448}      \\
Bckg ($\delta_{13} = 90^{\circ}$)    &416 &129   &297    &89       &403  &124   &286    &84     
\\
 \hline
 \end{tabular}}}
 \vspace {0.1in}
\caption{Total  neutrino ($N_{\alpha \beta}$) and anti-neutrino ($\bar N _{\alpha \beta}$) event rate at P2O corresponding to $\nu_{e}$ appearance and $\nu_{\mu}$ disappearance channels  for the $(3+0)$ and $(3+1)$ case.}
\label{table4_p2o}
\end{table}

It is clear that T2HK gives the best possibility to  distinguish between intrinsic and extrinsic $CP$ violating effects for $(3+0)$ case (see Eq.~\eqref{event_def}). For $(3+1)$ case, the corresponding quantity defined in Eq.~\eqref{event_def} is somewhat smaller.
For longer baselines, the matter contribution complicates the determination of the intrinsic $CP$ phase and it is expected that the ability of the other long baseline experiments will be limited~\cite{Rout:2017udo}. 

In what follows, we address the question of distinguishing the new physics scenario from the standard. 
\begin{figure}[t!]
\begin{center}
\includegraphics[width=6.5in]{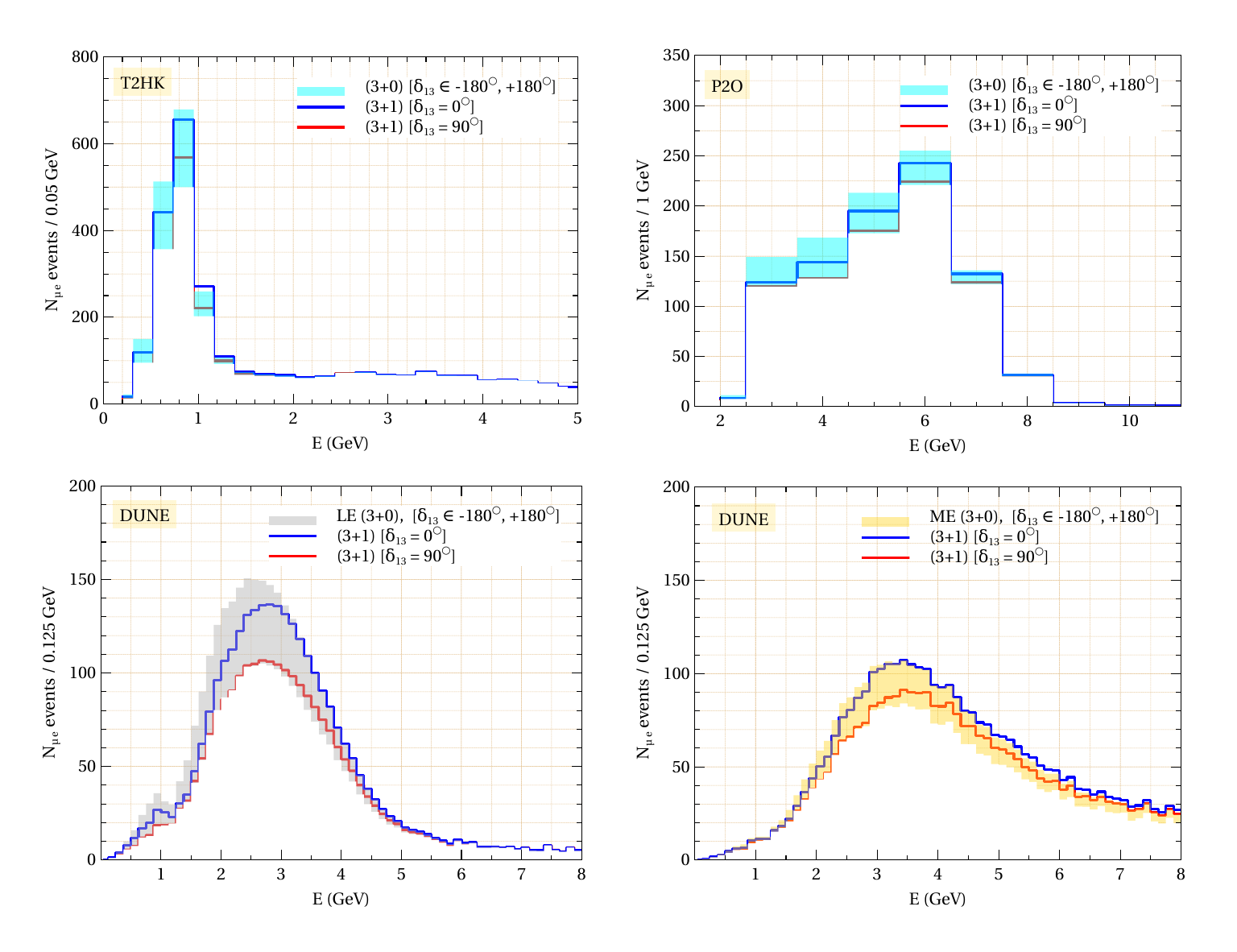}
\end{center}
\caption{Separation between $\nu_{\mu} \to \nu_{e}$ events  in the $(3+0)$  and $(3+1)$  cases for T2HK, DUNE and P2O. Upper row corresponds to $(3+0)$ events (cyan band) and $(3+1)$ events (blue and red lines 
for T2HK and P2O. Bottom row corresponds to DUNE, where we use the LE  beam (grey band) and ME beam (yellow band) for $\nu_{\mu} \to \nu_{e}$ channel. The cyan, grey and yellow band corresponds to the $(3+0)$ case with the full variation of $\delta_{13}$. For the $(3+1)$ case, we take fixed values of $\delta_{13}$ ($\delta_{13}=0^{\circ}$  shown in blue solid line and $\delta_{13}=90^{\circ}$  shown in solid red line). We take the values of sterile parameters as  $\Delta m^{2}_{41} = 1~\mbox{eV}^{2}, \theta_{14} = 5.7^{\circ},  \theta_{24} = 5^{\circ}, \theta_{34} = 20^{\circ}, \delta_{14} = \delta_{34} = 0^{\circ}$.}
\label{fig:event_def6}
\end{figure}
\begin{figure}[t!]
\begin{center}
\includegraphics[width=6.5in]{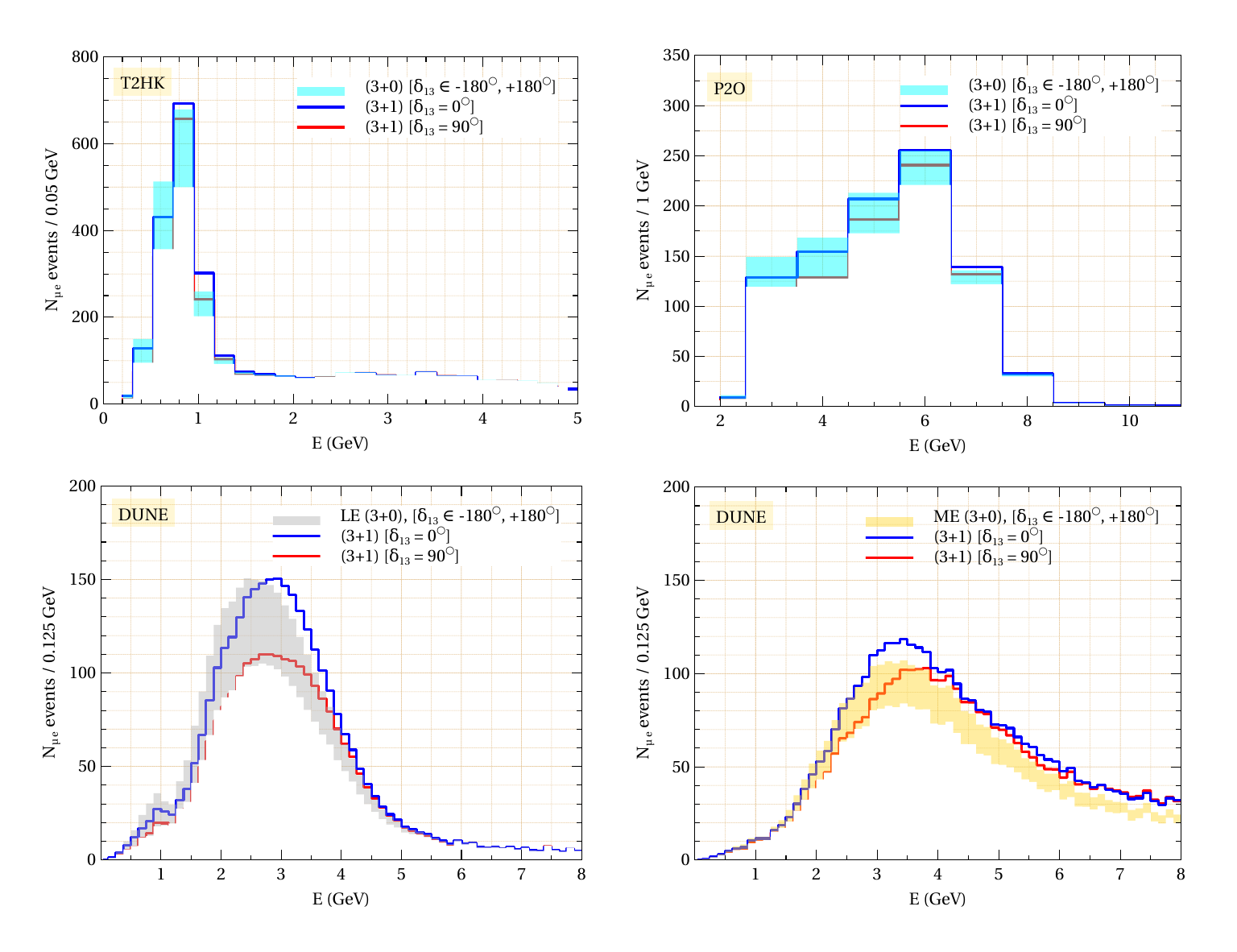}
\end{center}
\caption{Same as Fig.~\ref{fig:event_def6} but for different choice of sterile parameters. Here, we have taken $\Delta m^{2}_{41} = 1~\mbox{eV}^{2}, \theta_{14} = 18^{\circ},  \theta_{24} = 6^{\circ}, \theta_{34} = 25^{\circ}, \delta_{14} = \delta_{34} = 0^{\circ}$.}
\label{fig:event_def6n}
\end{figure}
\subsection{Distinguishing between the $(3+0)$ case and $(3+1)$ case}
\label{sec:chisq}
Let us now examine if the experiments considered in the present work would be able to distinguish between the standard $(3+0)$ case and the $(3+1)$ case. 
In order to assess if a given experiment would be able to distinguish between the two scenarios, we first obtain and compare the $\nu_{\mu} \to \nu_{e}$ event spectrum for the $(3+0)$ and $(3+1)$ cases.  In Fig.~\ref{fig:event_def6} and \ref{fig:event_def6n}, we plot the $\nu_{\mu} \to \nu_{e}$ event spectrum for the $(3+0)$ and $(3+1)$ cases corresponding to T2HK, DUNE and P2O. Note that we use two representative choices of sterile parameters - (a) using the true values ($\theta_{14} = 5.7^{\circ},  \theta_{24} = 5^{\circ}, \theta_{34} = 20^{\circ}$) and (b) using the values given by upper limit of the $3\,\sigma$ range ($\theta_{14} = 18^{\circ},  \theta_{24} = 6^{\circ}, \theta_{34} = 25^{\circ}$)  given in Table~\ref{parameters}. In both the figures, $\Delta m^{2}_{41} = 1~\mbox{eV}^{2}$ and the sterile phases are set to zero ($\delta_{14} = \delta_{34} = 0^{\circ}$).  If the sterile parameters are close to their true values, we find that the event spectra overlap (see Fig.~\ref{fig:event_def6}) while if larger values are taken as allowed by the $3\,\sigma$ range, the overlap of the event spectra is reduced thereby aiding the distinguishability of the two scenarios. We refer to these two choices as conservative and optimistic respectively. 
For the standard $(3+0)$ case, the cyan band corresponds to $\delta_{13} \in [-180^{\circ}, 180^{\circ}]$ for T2HK and P2O. In case of DUNE, we have two beam tunes, and the grey (yellow) band corresponds to $\delta_{13} \in [-180^{\circ}, 180^{\circ}]$ for LE (ME) beam.  For the $(3+1)$ case, we depict two cases, $\delta_{13} = 0^{\circ}$ in  blue and  $\delta_{13} = 90^{\circ}$   in red. As can be seen from Fig.~\ref{fig:event_def6}, the two curves lie within the cyan  band for T2HK or P2O (or grey or yellow band for DUNE) for conservative choice of parameters. There is some distinction possible between the $(3+1)$ curves and the ($3+0$) bands for the optimistic choice of sterile parameters (see  Fig.~\ref{fig:event_def6n}).
The events corresponding to $\delta_{13} = 0^{\circ}$ are shown in blue while the events corresponding to $\delta_{13} = 90^{\circ}$  are shown in red (see Fig.~\ref{fig:event_def6}). 

In order to distinguish between the $(3+0)$ and $(3+1)$ case, we define a metric~\footnote{The definition of the $\chi^{2}$ in
 Eq.~\ref{eq:chisq_si_str} includes only statistical effects and
  facilitates our understanding. The systematic effects are taken into
  account in the numerical results.} (see~\cite{Masud:2017bcf})
\bea
 \label{eq:chisq_si_str}
 \chi^{2}(\delta_{13}({true})) & = &
  \min_{\delta_{13}({test})} \sum_{i=1}^{x}  \sum_{j}^{2} 
  \frac{\bigg[N^{(3+1)}_{i,j}(\delta_{13}(true)) - N_{i,j}^{(3+0)} (\delta_{13}(test)\in[-180^{\circ}, 180^{\circ}])\bigg]^{2}}{N_{i,j}^{(3+1)} (\delta_{13}(true))}\nonumber
   \\
\eea
\begin{figure}[t!]
\includegraphics[width=6.9in]{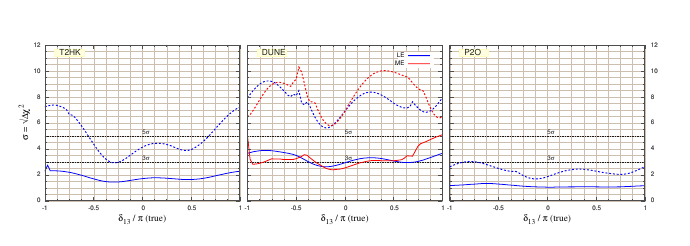}
\caption{Separation between $(3+0)$ and $(3+1)$ case for T2HK, DUNE (LE and ME) and P2O at $\chi^2$ level. We take two sets of values of sterile parameters : conservative choice  ($\Delta m^{2}_{41} = 1~\mbox{eV}^{2}, \theta_{14} = 5.7^{\circ},  \theta_{24} = 5^{\circ}, \theta_{34} = 20^{\circ}, \delta_{14} = \delta_{34} = 0^{\circ}$) shown as solid lines and optimistic choice ($\Delta m^{2}_{41} = 1~\mbox{eV}^{2}, \theta_{14} = 18^{\circ},  \theta_{24} = 6^{\circ}, \theta_{34} = 25^{\circ}, \delta_{14} = \delta_{34} = 0^{\circ}$) shown as dotted lines.}
\label{fig:chisq}
\end{figure}
\noindent
where, $N^{(3+0)}_{i,j}$ and $N^{(3+1)}_{i,j}$ are the test and true event datasets. We have marginalised over the  parameters, $\delta_{13}$, $\theta_{23}$ and $\Delta m^{2}_{31}$ in the test dataset as these are presently unknown. This $\chi^{2}$ was calculated using a set of conservative values of the sterile parameters ($\Delta m^{2}_{41} = 1~\mbox{eV}^{2}, \theta_{14} = 5.7^{\circ},  \theta_{24} = 5^{\circ}, \theta_{34} = 20^{\circ}, \delta_{14} = \delta_{34} = 0^{\circ}$)}. The index $i$ corresponds to energy bins ($i=1 \to x$, the number of bins depends on the particular experiment). For T2HK, there are $x=95$ bins of width $0.05$ GeV in $0.1$ - $5$ GeV~\cite{Hyper-Kamiokande:2018ofw}. For DUNE, 
there are $x=62$ energy bins each having a width of $0.125$ GeV in the energy range of $0.5$ - $8$ GeV and $2$ bins of width 1 GeV each in the range $8$ - $10$ GeV~\cite{DUNE:2021cuw} and for P2O, there are $x=10$ bins of width $1$ GeV in $1.5$ - $11.5$ GeV)~\cite{Akindinov:2019flp}. The index $j$ is summed over the modes ($\nu$ and $\bar{\nu}$) respectively.

In Fig.~\ref{fig:chisq}, we plot the quantity defined above as a function of $\delta_{13}$ for the three experiments - T2HK, DUNE  and P2O. It can be noted that for conservative choice (i.e., values close to the true values given in Table~\ref{parameters}) of sterile parameters, only DUNE (with LE and ME beam tune) will be able to distinguish between the scenarios with sensitivity above $3\,\sigma$. Whereas for the other two experiments (T2HK and P2O), the sensitivity is below $3\,\sigma$. The reason behind this can be understood as follows. The sensitivity of P2O is limited because it has poor background rejection capability~\cite{Akindinov:2019flp}.
If we use optimistic choice (i.e., upper limit of the $3\,\sigma$ range  given in Table~\ref{parameters}) of the sterile parameters, we note that  the ability to distinguish between the scenarios in general  improves for the all the three experiments. For T2HK, the sensitivity lies above $3\,\sigma$ for all values of $\delta_{13}$. For P2O, the sensitivity remains below $3\,\sigma$ for all values of $\delta_{13}$. For DUNE (with LE and ME beam tune), we will be able to distinguish between the scenarios with sensitivity above $5\,\sigma$. 
{\subsection{Expected precision of active-sterile mixing angles}
\label{sec:precision}
In order to address the question of precision attainable at future long baseline neutrino experiments, we can define the  
 $\chisq$ as~\cite{Fiza:2021gvq}
 \begin{align}
\label{eq:chisq}
\Delta \chi^{2}(p^{\text{true}}) = \underset{p^{\text{test}}, \eta}{\text{Min}} \Bigg[&2\sum_{k}^{\text{mode}}\sum_{j}^{\text{channel}}\sum_{i}^{\text{bin}}\Bigg\{
N_{ijk}^{\text{test}}(p^{\text{test}}; \eta) - N_{ijk}^{\text{true}}(p^{\text{true}})\nonumber\\
&+ N_{ijk}^{\text{true}}(p^{\text{true}}) \ln\frac{N_{ijk}^{\text{true}}(p^{\text{true}})}{N_{ijk}^{\text{test}}(p^{\text{test}}; \eta)} \Bigg\} 
+ \sum_{l}\frac{(p^{\text{true}}_{l}-p^{\text{test}}_{l})^{2}}{\sigma_{p_{l}}^{2}}
+ \sum_{m}\frac{\eta_{m}^{2}}{\sigma_{\eta_{m}}^{2}}\Bigg],
\end{align}
where $N^{\text{true}}$ ($N^{\text{test}}$) denote the event rates for {\it{true}}  ({\it{test}}) datasets, while $p^{\text{true}}$ ($p^{\text{test}}$) represent the set of {\it{true}} ({\it{test}}) oscillation parameters. 
The index $i$ is summed over the energy bins of the experiments. 
The indices $j$ and $k$ account for the oscillation channels ($\nu_e, \nu_\mu, \nu_\tau$) including NC channels and the modes ($\nu$ and $\bar{\nu}$), respectively.  
The term $(N^{\text{test}} - N^{\text{true}})$ accounts for the algebraic difference while the logarithmic term enclosed within curly braces quantifies the fractional difference of the two datasets. 
Together, the summation over $i, j, k$ within the curly braces forms the statistical component of the $\chisq$ function.
The values of the true or best-fit oscillation parameters and their uncertainties  used in the present analysis are tabulated in Table~\ref{parameters}. The prior uncertainty of the $l^{{th}}$  oscillation parameter, $p_l$, is represented by $\sigma_{p_l}$.  
In the last term, $\sigma_{\eta_{m}}$ is the uncertainty on the systematic/nuisance parameter $\eta_{m}$ and 
the sum over $m$ takes care of the systematic part of $\chisq$. 
This way of treating the systematics in the $\chisq$ calculation is known as the {\it{method of pulls}}~\cite{Huber:2002mx,Fogli:2002pt,GonzalezGarcia:2004wg,Gandhi:2007td}. 
}  

{{In Fig.~\ref{fig:chisq_precision}, we demonstrate how efficiently a given long baseline experiment will be able to reconstruct the CP phase in correlation with active-sterile mixing angle at $1\,\sigma$ C.L. The two rows correspond to the expected contours for different experiments whereas the three columns correspond to the three different active-sterile mixing angles. For certain choice of parameters we obtain closed contours from which we can deduce the expected precision on those parameters. In Table~\ref{tab:chisq_constraints}, we list the expected precision on active-sterile mixing angles ($\theta_{14}$, $\theta_{24}$ and $\theta_{34}$) for conservative and optimistic choice of parameters at DUNE and T2HK. For $\theta_{34}$, the contours are not closed for either of the choices, resulting in lower limit only. For $\theta_{14}$, the contours are not closed for the optimistic choice, resulting in lower limit only.  It may be noted that it is difficult to constrain any of the active-sterile mixing angle using P2O. }}

\begin{figure}[t]
\includegraphics[width=6.9in]{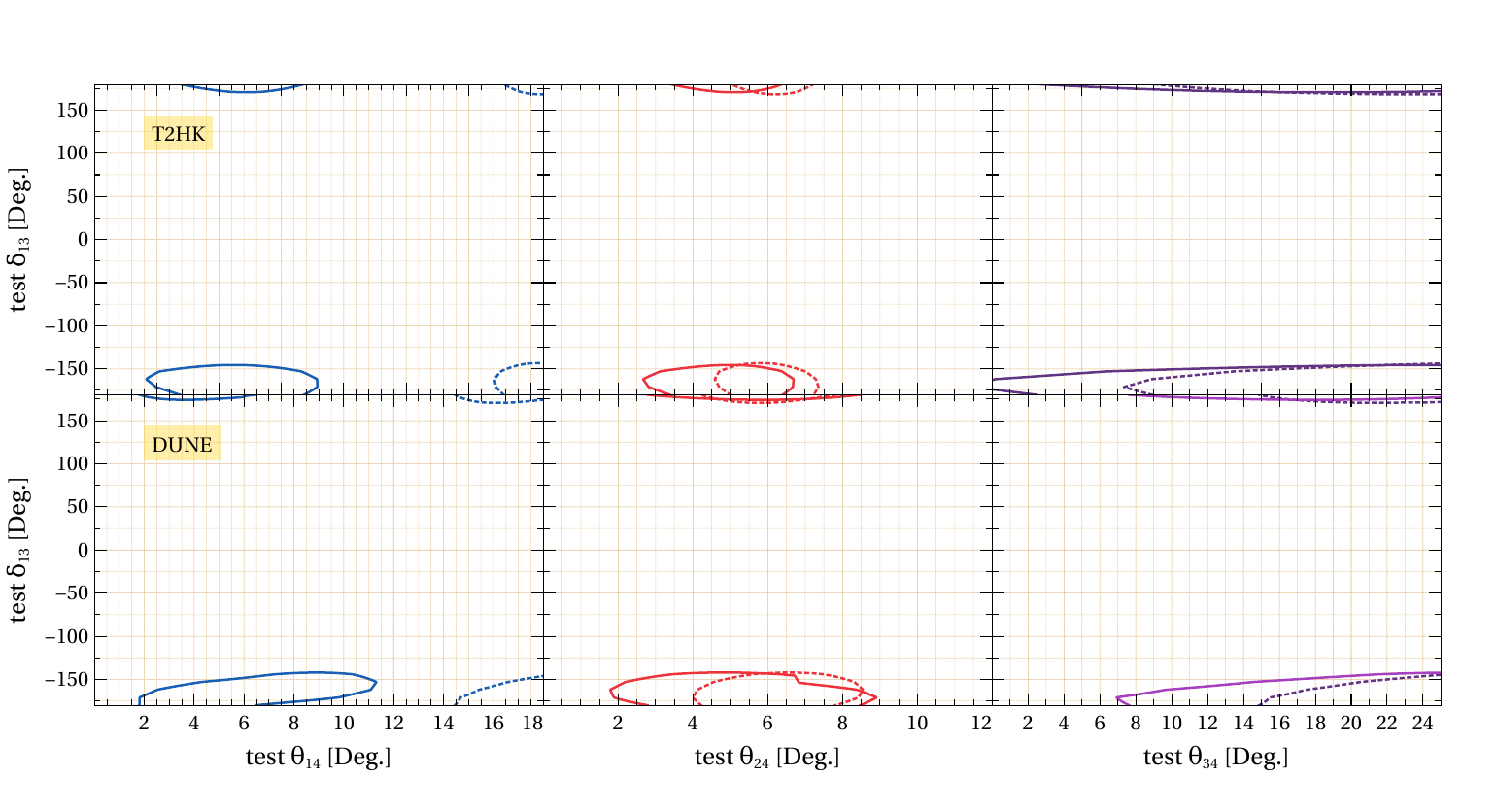}
\caption{The figure shows the $\Delta \chi^{2}$ contours to assess the precision of sterile parameters at $1\,\sigma$ C.L. in the plane of $(\theta_{i4}-\delta_{13})$ where $i=1,2,3$ for T2HK (upper row) and DUNE (lower row). We take two sets of values of sterile parameters : conservative choice ($\Delta m^{2}_{41} = 1~\mbox{eV}^{2}, \theta_{14} = 5.7^{\circ},  \theta_{24} = 5^{\circ}, \theta_{34} = 20^{\circ}, \delta_{14} = \delta_{34} = 0^{\circ}$) shown as solid contours and optimistic choice ($\Delta m^{2}_{41} = 1~\mbox{eV}^{2}, \theta_{14} = 18^{\circ},  \theta_{24} = 6^{\circ}, \theta_{34} = 25^{\circ}, \delta_{14} = \delta_{34} = 0^{\circ}$) shown as dotted contours.}
\label{fig:chisq_precision}
\end{figure}
\noindent
\begin{table}[t]
    \centering
{\small{    \begin{tabular}{|c|c|c|c|}
    \hline
& \multicolumn{3}{c|}{Conservative choice } \\ 
    \cline{2-4}
        T2HK  & $2.1^{\circ}\leq\theta_{14}^{}\leq8.8^{\circ}$ & $2.6^{\circ}\leq \theta_{24}^{}\leq 6.8^{\circ}$ & $-$ \\  \cline{2-4}
        DUNE & $2.0^{\circ}\leq\theta_{14}^{}\leq11.5^{\circ}$ & $1.8^{\circ}\leq\theta_{24}^{}\leq8 .8^{\circ}$ & $>7.0^{\circ}$ \\ \hline
 & \multicolumn{3}{c|}{Optimistic choice} \\ 
       \cline{2-4}
        T2HK  & $>16.0^{\circ}$ & $4.5^{\circ}\leq\theta_{24}^{}\leq 7.2^{\circ}$ & $>7.3^{\circ}$ \\  \cline{2-4}
        DUNE  & $>14.5^{\circ}$ & $4.0^{\circ}\leq\theta_{24}^{}\leq 8.4^{\circ}$ & $>15.0^{\circ}$ \\ \hline
    \end{tabular}}}
    \caption{\footnotesize{Reconstructed range of $\theta_{i4}$ ($i=1,2,3$) for conservative choice ($\theta_{14}=5.7^{\circ}$, $\theta_{24}=5^{\circ}$, $\theta_{34}=20^{\circ}$) and optimistic choice  ($\theta_{14}=18^{\circ}$, $\theta_{24}=6^{\circ}$, $\theta_{34}=25^{\circ}$)  of sterile parameters for T2HK and DUNE. }}
    \label{tab:chisq_constraints}
\end{table}

\section{Conclusion}
\label{sec:conclusion}

One of the well-motivated and widely studied new physics scenarios is that of an eV-scale sterile neutrino mixing with the three active neutrinos of the standard paradigm (for instance, see reviews~\cite{Giunti:2019aiy, Dasgupta:2021ies}).  In the present work, we explore how an eV-scale sterile neutrino impacts the measurements at long baseline experiments. In particular, we focus on (a) $CP$  violating probability differences for the different appearance and disappearance  channels  and discuss the role played by different channels, (b) the question of cleanly identifying the contribution to $CP$ violation coming from intrinsic source in presence of matter, and finally (c) whether long baseline experiments have the ability to distinguish between the $(3+0)$ and $(3+1)$ case.

We carry out detailed numerical simulations for three upcoming long baseline experiments which employ different detection techniques. T2HK and P2O have a WC detector while DUNE is based on the LArTPC technology. The ability of T2HK to distinguish between intrinsic and extrinsic effects in the  $(3+0)$ case and $(3+1)$ case is shown in (see Fig.~\ref{fig:event_def1} and Table~\ref{table1_t2hk}). The shorter baseline allows for a cleaner separation between intrinsic and extrinsic $CP$ effects.

For the case of  DUNE, we consider two beam tunes (LE and ME) and  explore the possibility to distinguish between intrinsic and extrinsic effects in the  $(3+0)$ case as well as the $(3+1)$ case (see Fig.~\ref{fig:event_def2} and Fig.~\ref{fig:event_def3}, Table~\ref{table2_LE} and Table~\ref{table3_ME}). The ME flux almost does not impact the $\nu_\mu \to \nu_e$  channel but has a role in the $\nu_\mu \to \nu_\mu$ channel. P2O offers much less sensitivity as can be seen from  Fig.~\ref{fig:event_def4} and Table~\ref{table4_p2o}.

Finally, in Sec.~\ref{sec:chisq}, we address the question of separating between standard $(3+0)$ case and new physics $(3+1)$ scenario. For the $\nu_\mu \to \nu_e$ channel, we examine the event rates for the different experiments to assess if a given experiment can allow us to distinguish between  $(3+0)$ case and $(3+1)$ case (see Fig.~\ref{fig:event_def6} and Fig.~\ref{fig:event_def6n}).
Then, we perform a detailed $\chi^{2}$ analysis (using all channels) for the considered experiments (see Fig.~\ref{fig:chisq}). 

From Fig.~\ref{fig:chisq}, it can be seen that DUNE has $> 3\,\sigma$ sensitivity for all values of $\delta_{13}$ to distinguish between the two scenarios  even for the conservative choice of sterile parameters \ie, when the sterile parameters are taken to be true values given in Table~\ref{parameters}. For optimistic scenario \ie, larger allowed values of the sterile parameters (given by upper limit of the $3\,\sigma$ range in Table~\ref{parameters}), we note that DUNE has $> 5\,\sigma$ sensitivity for all values of $\delta_{13}$ to distinguish between the two scenarios. As far as T2HK is concerned, only for the optimistic choice of sterile parameters, we obtain $> 3\,\sigma$ sensitivity for all values of $\delta_{13}$ to distinguish between the two scenarios. It is clear that P2O will have much less ($< 3\,\sigma$) sensitivity for all values of $\delta_{13}$ to distinguish between the two scenarios.

{{Finally, we also deduce the expected precision on the active-sterile mixing angles at future long baseline experiments such as T2HK and DUNE. Fig.~\ref{fig:chisq_precision} and Table~\ref{tab:chisq_constraints} summarize our results on the expected precision on active-sterile mixing for the two choices of sterile parameters considered in the present work.}}

{\textbf{Acknowledgment :}} {SP acknowledges Jogesh Rout and Mehedi Masud for useful discussions related to the GLoBES software. KS would like to thank PM for the warm hospitality, fruitful collaboration and the useful discussions on neutrino oscillations during her visits to JNU, New Delhi. PM would like to thank Mary Bishai for many valuable discussions and for providing the medium energy flux files used for analysis in the present work. SP acknowledges JNU for support in the form of fellowship. The  numerical analysis has been performed using the HPC cluster at SPS, JNU funded by DST-FIST}. KS acknowledges financial support from the Ministry of Education, Government of India. It is to be noted that this work has been done by the authors and is not representation of the DUNE collaboration. This research (SP and PM) was supported in part by the International Centre for
Theoretical Sciences (ICTS) for participating in the program - Understanding the Universe Through Neutrinos (code: ICTS/Neus2024/04).
\appendix
\renewcommand{\theequation}{\thesection.\arabic{equation}}
\setcounter{equation}{0}
\counterwithin{figure}{section}
\numberwithin{equation}{section}
\renewcommand{\thesection}{\Alph{section}}
\renewcommand{\thesubsection}{\Alph{subsection}}
\section{Oscillation probabilities in the $(3+1)$ case}
 \label{app_1}

For the $(3+0)$ case, we use the expressions obtained in~\cite{Akhmedov:2004ny}. In this section, we give the approximate expressions for oscillation proabilities in the $(3+1)$ scenario. We adopt the parameterization for the mixing matrix as given in Eq.~\eqref{eq:umat}.
Following the procedure laid down in~\cite{Klop:2014ima}, we obtain the probability expressions in the $(3+1)$ case for  different appearance and disappearance channels. In what follows, the small parameters are identified as follows:  $s_{13}$, $s_{14}$ and $s_{24}$ (which are almost equal) are  treated to be ${\cal O}(\epsilon)$ where $\epsilon$ is a small parameter which is ${\cal O}(10^{-1})$ (for justification, see $(3+0)$ global-fit results~\cite{Esteban:2016qun,PhysRevD.101.116013} and $(3+1)$ fit results~\cite{Giunti:2013aea, Kopp:2013vaa,Dentler:2018sju}). The hierarchy parameter, $\alpha = \Delta m^2_{21}/ \Delta m^2_{31}$ has a value $\simeq 0.03$ and can be taken to be ${\cal O}(\epsilon^2)$. The oscillations induced by $\Delta m^{2}_{41}\sim 1~\text{eV}^{2}$ have been averaged out.

{\underline{\bf{Appearance channels}}}
 
 $P^{(3+1)}_{\mu e} $ is a sum of three contributions~\cite{Klop:2014ima},
\bea
P^{(3+1)}_{\mu e} &\simeq& (1+2\hat{A})P^{\rm {ATM}}+ P^{\rm {INT}}_{\rm I} + P^{\rm {INT}}_{\rm II}, 
\quad \textrm{with} \quad \nonumber
\\
P^{\rm {ATM}} &\simeq& 4 s_{23}^2 s^2_{13}  \sin^2{\Delta}\, ,\nonumber \\
 P^{\rm {INT}}_{\rm I} &\simeq& 8 s_{13} s_{12} c_{12} s_{23} c_{23} (\alpha \Delta)\sin \Delta \cos({\Delta + \delta_{13}})\,,\nonumber
\\
P^{\rm {INT}}_{\rm II} &\simeq& 4 s_{14} s_{24} s_{13} s_{23} \sin\Delta \sin (\Delta + \delta_{13} - \delta_{14})\,.
\eea
Here, 
$ \Delta = {\Delta m^{2}_{31} L}/{4E},\,\alpha = \Delta m^2_{21}/ \Delta m^2_{31},\,\hat A = {A}/{{\Delta m^{2}_{31}}},\,A = 2 \sqrt{2} G_F N_e E$.
$ P^{\rm {ATM}}$ is the ${\cal O} (\epsilon^2)$ contribution from atmospheric sector, $P^{\rm {INT}}_{\rm I}$ is the ${\cal O} (\epsilon^3)$ contribution from the solar-atmospheric interference and $P^{\rm {INT}}_{\rm II}$ is the ${\cal O} (\epsilon^3)$ contribution from  atmospheric-sterile interference.
For $\nu_{\mu} \to \nu_{\tau}$, transition probability is given by
\bea
P^{(3+1)}_{\mu\tau} &\simeq& c^2_{24}c^2_{34} P^{(3+0)}_{\mu \tau} + P^{\rm {INT}}_{\rm I} + P^{\rm {INT}}_{\rm II} + P^{\rm {INT}}_{\rm III}\,,  \quad \textrm{with} \quad \nonumber \\
 P^{(3+0)}_{\mu \tau} &\simeq& \sin^2 2\theta_{23} \sin^2 \Delta -\alpha c^2_{12}  \sin^2 2\theta_{23}\Delta \sin 2\Delta +\alpha^2 c^4_{12}  \sin^2 2\theta_{23}\Delta^2 \cos 2\Delta \, \nonumber \\
&+& \frac{2}{\hat{A}-1}s^2_{13}\sin^2 2\theta_{23}\bigg[\sin \hat{A}\cos \hat{A}\Delta \frac{\sin(\hat{A}-1)\Delta}{\hat{A}-1}-\frac{\hat{A}}{2}\Delta \sin 2\Delta\bigg] \nonumber \\
&+& 2\alpha s_{13} \sin 2\theta_{12}\sin 2\theta_{23}\sin \delta_{13}\sin \Delta \frac{\sin \hat{A}\Delta}{\hat{A}}\frac{\sin(\hat{A}-1)\Delta}{\hat{A}-1} \, \nonumber \\
&-& \frac{2}{\hat{A}-1}\alpha s_{13} \sin 2\theta_{12}\sin 2\theta_{23} \cos 2\theta_{23}\cos \delta_{13}\sin \Delta \Big[\hat{A}\sin\Delta -\frac{\sin \hat{A}\Delta}{\hat{A}} \cos(\hat{A}-1)\Delta\Big] \,, \nonumber\\
P^{\rm {INT}}_{\rm I} &\simeq& \frac{-2}{\hat{A}-1}s_{14}s_{24}c_{24}c^2_{34}\bigg\{\cos(\delta_{14}-\delta_{13})c_{23}+\cos(\Delta-\delta_{13}/2)\cos\big(\delta_{14} + \delta_{13}/2+\omega_{-} \big) \cos 2\theta_{23} \nonumber \\ &-&\sin(\delta_{14}-\delta_{13}/2)\sin\big(\Delta + \delta_{13}/2+ \omega_{-} \big) \sin \theta_{13}\bigg\} \,,\nonumber \\
P^{\rm {INT}}_{\rm II} &\simeq& \frac{-2}{\hat{A}-1}c_{24}s_{14}s_{34}c_{34}c^2_{24}\cos\big[\delta_{13}-\delta_{14}-2(\hat{A}-1)\Delta\big]s_{13} s_{23} \,,\nonumber \\
P^{\rm {INT}}_{\rm III}&\simeq&-2s_{24}s_{34}c_{34}c^2_{24}\cos(\Delta -\delta_{13}/2)\cos\big[\delta_{34}+ \omega_{-} -\delta_{13}/2 \big]\cos 2\theta_{23} \nonumber \\
&-&\sin(\Delta -\delta_{13}/2) \sin\big[\delta_{34}+ \omega_{-} -\delta_{13}/2 \big]\,.
\eea
where we have considered terms upto ${\cal O} (\epsilon^3)$. Here,
\bea
\Omega_{\pm} &=&  \frac{\cos(\Delta \pm \delta_{13}/2)}{\hat A \pm 1},\,\nonumber \Phi_{\pm} = \frac{\sin(\Delta \pm \delta_{13}/2)}{\hat A \pm 1},\,\,\omega_{\pm} = (1 \pm 2 \hat A ) \Delta\,. \nonumber
\eea
In this case, the leading contribution comes from the probability for the $(3+0)$ case with some factors. $P^{\rm {INT}}_{\rm I}$, $P^{\rm {INT}}_{\rm II}$ and $P^{\rm {INT}}_{\rm III}$ are ${\cal O} (\epsilon^2)$, ${\cal O} (\epsilon^3)$ and  ${\cal O} (\epsilon^3)$ contributions respectively from  active-sterile interference.

For $\nu_{e} \to \nu_{\tau}$ transition, the probability is given by
\bea
P^{(3+1)}_{e\tau} &\simeq& c^2_{14}c^2_{24}P^{(3+0)}_{e\tau} - \frac{2}{\hat{A}-1}c^2_{14}c_{24}c_{34}s_{14}s_{34}\bigg\{\cos(\delta_{14}-2\delta_{13})\cos\theta_{23}+\cos(\Delta-\delta_{13}/2) \, \nonumber \\
&&\cos\big(\delta_{14}-\delta_{13}/2+\omega_{-}\big)\cos2\theta_{23}-\sin(\Delta-\delta_{13}/2)\sin\big(\delta_{14}-\delta_{13}/2+\omega_{-})s_{13}\bigg\} \,,  \nonumber 
\\ \nonumber
\textrm{with}
\\
P^{(3+0)}_{e\tau} &\simeq& \alpha ^2 \sin^2 2\theta_{12}s^2_{23}\frac{\sin ^2 A\Delta}{A^2} + 4 s^2_{13}c^2_{23}\frac{\sin ^2(A-1)\Delta}{(A-1)^2} \, \nonumber \\
&-& 2\alpha_{13}\sin 2\theta_{12} \sin 2\theta_{23}\cos (\Delta-\delta_{13})\frac{\sin A\Delta}{A}\frac{\sin(A-1)\Delta}{(A-1)}\, .
\eea
where we have considered terms upto ${\cal O} (\epsilon^3)$. In this case, the leading contribution comes from the probability for the $(3+0)$ case with some factors.

 {\underline{\bf{Disppearance channels}}}
 
The $\nu_e \to \nu_e$ disappearance probability at ${\cal O} (\epsilon^3)$ is
\bea
P^{(3+1)}_{ee} &\simeq&(1-2 s^2_{14})P^{(3+0)}_{ee} \,,\quad \textrm{with} \quad 
\nonumber \\
P^{(3+0)}_{ee} &\simeq& 1-\alpha^2 \sin^2 2\theta_{12}\frac{\sin^2 A\Delta}{A^2}-4s^2_{13}\frac{\sin^2(A-1)\Delta}{(A-1)^2}\,.
\eea
The expression for $\nu_\mu \to \nu_\mu$ disappearance probability is~\cite{Sharma:2022zaf}. 
\bea
 P^{(3+1)}_{\mu \mu} &\simeq&  P^{(3+0)}_{\mu \mu}  + P^{\rm I}_{\rm INT} + P^{\rm II}_{\rm INT} + P^{\rm III}_{\rm INT} \,, \quad \textrm{with} \quad \nonumber
\\
 P^{(3+0)}_{\mu \mu} &\simeq& 1 - \sin^22\theta_{23} \sin^2{\Delta} + \alpha c^2_{12} \sin^22\theta_{23} \Delta \sin2\Delta  - 4 s^2_{13} s^2_{23} \frac{\sin^2(\hat{A} - 1)\Delta}{(\hat{A} - 1)^2}  \, \nonumber \\
&-& \frac{2}{\hat{A} - 1} s^2_{13}\sin^22\theta_{23}  \Big[\sin\Delta \cos\hat{A}\Delta \frac{\sin(\hat{A} - 1)\Delta}{\hat{A} - 1} - \frac{\hat{A}}{2} \Delta \sin2\Delta \Big] \, \nonumber \\
&-& 2 \alpha s_{13} \sin2\theta_{12} \sin2\theta_{23} \cos\delta_{13} \cos\Delta \frac{\sin\hat{A}\Delta}{\hat{A}} \frac{\sin(\hat{A} - 1)\Delta}{\hat{A} - 1}\, \nonumber \\
&+& \frac{2}{\hat{A} - 1} \alpha s_{13} \sin2\theta_{12} \sin2\theta_{23} \cos2\theta_{23} \cos\delta_{13} \sin\Delta  \Big[\hat{A} \sin\Delta - \frac{\sin\hat{A}\Delta}{\hat{A}} \cos(\hat{A} - 1) \Delta \Big]\,,\nonumber
 \\
 P^{\rm I}_{\rm INT} &\simeq& -2 s^2_{14}\bigg[ 1 - \sin^22\theta_{23} \sin^2{\Delta} \bigg] \,, \nonumber
\\
 P^{\rm II}_{\rm INT} &\simeq&
\frac{1}{4(\hat{A} -1)} s_{13} c_{24} s_{14} s_{24} \bigg\{\bigg(-3\cos \big(\delta_{14} + (\hat{A} -1)\Delta\big) + \cos \big(\delta_{14} + 2 \hat{A}\Delta +\delta_{13}\big)\bigg)s_{23} \, \nonumber \\
 &-& 2 \sin(\delta_{14}+\delta_{13})\sin \big(2\Delta - \delta_{13}\big) \sin 2 \theta_{23}\, +2\cos \big(2\Delta-\delta_{13}\big) 
\bigg(\cos \Big[\delta_{14} -\omega_{-} -\delta_{13}/2 \Big]
\sin 3\theta_{23} \, \nonumber \\
& +& \cos \big(\Delta- \delta_{13}/2 \big) \cos (\delta_{14}+\delta_{13}) \sin 4\theta_{23} \bigg) \bigg\} \,,\nonumber
\\
 P^{\rm III}_{\rm INT} &\simeq&
\frac{1}{2(\hat{A} -1)} s_{14} s_{13} s_{23} s_{24} c_{24}\bigg\{\cos (\delta_{14} - 2\Delta) - \cos(\delta_{14}-\delta_{13}) + 2 \cos \big(\Delta - \delta_{13}/2  \big) \nonumber \\ && \cos \big(\delta_{14}-\Delta- \delta_{13}/2  \big)\cos 2\theta_{23}\bigg\} \,.
\eea
where, we have considered terms upto ${\cal O} (\epsilon^3)$. In this case, the leading contribution comes from the probability for the $(3+0)$ case with some factors. $P^{\rm {INT}}_{\rm I}$, $P^{\rm {INT}}_{\rm II}$ and $P^{\rm {INT}}_{\rm III}$ are ${\cal O} (\epsilon^2)$, ${\cal O} (\epsilon^3)$ and ${\cal O} (\epsilon^3)$ contributions from  active-sterile interference.
 
The expression for $\nu_\tau \to \nu_\tau$ disappearance probability upto  ${\cal O} (\epsilon^3)$ is
\bea
P^{(3+1)}_{\tau \tau} 
&\simeq& \,c_{34}^2 P^{(3+0)}_{\tau \tau} + P^{INT}_{I}\,, \quad \textrm{with} \quad \nonumber 
\\
P^{(3+0)}_{\tau \tau} &\simeq& 1 - (P^{(3+0)}_{e \tau} + P^{(3+0)}_{\mu \tau}) \,,\nonumber 
\\
 P^{INT}_{I} &\simeq& \frac{s_{13}c_{34}^3}{16(\hat{A}-1)}\bigg\{6\cos(\delta_{14}-3\delta_{13}) -4\cos(\delta_{14}-2\Delta-2\delta_{13})+6\cos(\delta_{14}-4\Delta-\delta_{13}) \,\nonumber \\
 &+& \cos(\delta_{14}-3\delta_{13}-4\theta_{23})+\cos(\delta_{14}-4\Delta-\delta_{13}-4\theta_{23}) + 12\cos\big(\delta_{14}-2(1+\hat{A})\Delta)c_{23}\big)  \nonumber \\
 &-& 4\cos(\delta_{14}-2\hat{A}\Delta-\delta_{13})c_{23}+8\cos(\Delta -\delta_{13})\cos(-\delta_{14}+\omega_{+}+\delta_{13}/2)\cos3\theta_{23} \nonumber \\
 &+& 4\cos^2(\Delta-\delta_{13}/2)\cos(\delta_{14}-2\Delta-2\delta_{13}+4\theta_{23})+2\cos(\delta_{14}-2\Delta-2\delta_{13}-4\theta_{23}) \nonumber \\
 &+& 16\cos2\theta_{23}\sin(\delta_{14}-2\Delta-2\delta_{13})\sin(2\Delta-\delta_{13}) \bigg\}\, .
\eea
 In this case, the leading contribution comes from the probability for the $(3+0)$ case with some factors. $P^{\rm {INT}}_{\rm I}$ is ${\cal O} (\epsilon)$ contribution from  active-sterile interference. There are additional terms upto ${\cal O} (\epsilon^3)$ at the sub-leading order but their contributions to the overall probability can be neglected safely.
\bibliographystyle{unsrt}
\bibliography{reference}
\end{document}